\documentclass[preprint,3p,onecolumn,numbers]{elsarticle}

\usepackage{amssymb}
\usepackage{amsmath}
\usepackage{dsfont}
\usepackage{bm,bbm}
\usepackage{lipsum}
\usepackage{mathtools, cuted}
\usepackage{breqn}
\usepackage{xcolor}
\usepackage{soul}
\usepackage{caption}
\usepackage{subcaption}
\usepackage{graphicx}
\usepackage{amsthm}
\usepackage{verbatim}
\usepackage{dsfont}
\usepackage{hyperref}
\usepackage{enumerate}

\usepackage{xcolor}

\definecolor{mygreen}{RGB}{51, 160, 44} % Matplotlib green
\definecolor{mypurple}{RGB}{152, 78, 163} % Matplotlib purple

\usepackage{amsmath}

\usepackage[capitalise]{cleveref}

\begin{document}

\begin{frontmatter}

\title{Financial instability transition under heterogeneous investments and portfolio diversification
}

\author[1]{Preben Forer}
\author[2]{Barak Budnick}
\author[1]{Pierpaolo Vivo}
\author[3]{Sabrina Aufiero}
\author[3]{Silvia Bartolucci}
\author[3,4,5]{Fabio Caccioli}

\affiliation[1]{organization={Department of Mathematics, King’s College London}, addressline={Strand}, city={London}, postcode={WC2R 2LS}, country={United Kingdom}}

\affiliation[2]{organization={Racah Institute of Physics, The Hebrew University}, city={Jerusalem}, postcode={9190401}, country={Israel}}

\affiliation[3]{organization={Department of Computer Science, University College London}, addressline={66-72 Gower Street}, city={London},postcode={WC1E 6EA}, country={United Kingdom}}

\affiliation[4]{organization={Systemic Risk Centre, London School of Economics and Political Sciences}, city={London}, postcode={WC2A 2AE}, country={United Kingdom}}
            
\affiliation[5]{organization={London Mathematical Laboratory}, addressline={8 Margravine Gardens}, city={London}, postcode={WC 8RH}, country={United Kingdom}}

\begin{abstract}
We analyze the stability of financial investment networks, where financial institutions hold overlapping portfolios of assets. We consider the effect of portfolio diversification and heterogeneous investments using a random matrix dynamical model driven by portfolio rebalancing. While heterogeneity generally correlates with heightened volatility, increasing diversification may have a stabilizing or destabilizing effect depending on the connectivity level of the network. The stability/instability transition is dictated by the largest eigenvalue of the random matrix governing the time evolution of the endogenous components of the returns, for which different approximation schemes are proposed and tested against numerical diagonalization.

\end{abstract}

\end{frontmatter}
\tableofcontents

\newpage

\section{Introduction}

Financial contagion, the spread of shocks through interconnected financial institutions, operates primarily through two main channels: direct and indirect contagion. These channels are underpinned by the dual roles banks play within the financial system -- as lenders to each other and as investors in shared asset markets. Understanding these contagion routes is essential for analyzing the stability of financial networks and the broader economic ecosystem.

Direct contagion occurs when financial distress at one bank directly impacts the creditors linked to it. For instance, if a bank defaults on its obligations, this can compromise the financial health of other institutions that rely on those repayments. This one-to-one transmission of financial stress can propagate through the inter-bank network, potentially leading to a cascade of bank insolvencies. Research in network science, pioneered by \cite{Gai2010}, has laid the groundwork for modeling these direct contagion pathways, providing insights into how financial networks respond to shocks and identifying strategies to limit contagion, such as applying transaction taxes to reduce systemic risk or implementing targeted risk immunization policies \cite{Caccioli2011}.

Indirect contagion, in contrast, arises through the interaction of banks operating in asset markets. Through this route, financial pressure on one bank can indirectly affect others through shared asset holdings. For example, if multiple banks hold significant positions in the same stock, a forced sale by one institution to meet liquidity demands may drive down the asset’s price. This price drop, in turn, decreases the value of the same asset held by other banks, prompting further asset sales and potentially initiating a fire sale cascade. This form of contagion, driven by overlapping portfolios, has been identified as a significant factor in the 2008 financial crisis  \cite{Acharya2009}.

Studying indirect contagion presents unique challenges. Unlike the direct contagion path, which involves modeling a chain reaction of direct solvency shocks, indirect contagion requires an examination of asset price dynamics. Banks no longer directly impact each other, but now do so through an intermediary: the common asset owned \cite{Huang2013,Caccioli2012b}. To address this, models have been developed to represent the bank-asset system as a bipartite network, linking banks to the assets they hold. There are two main drivers of indirect contagion: overlaps among banks’ investment portfolios, and the manner in which banks manage their risk/exposure. Often, exposure management is what triggers contagion, while the portfolio overlaps are what enables it to spread.

In traditional economic models, financial institutions are often viewed as passive entities whose risk management practices have limited influence on broader economic stability. However, empirical research reveals that the collective growth of credit by institutions frequently precedes periods of financial instability \cite{Gourinchas2001, Mendoza2012, Borio2009}. This suggests that the role of institutional exposure management in economic cycles is more active and influential than traditionally believed. Rather than merely reflecting external economic changes, the balance sheet constraints of financial institutions actively contribute to the cycles of economic growth and recession. They force financial institutions to transact, which can then trigger indirect financial contagion \cite{Reinhart2009, Schularick2012}.

This paper aims to explore how certain banking practices -- such as portfolio diversification (spreading investments across various assets) and heterogeneous investment strategies (making investments of different sizes) -- influence the amount of financial instability present through indirect contagion. While diversification theoretically reduces risk by balancing potential losses across different assets, it also leads to shared or “overlapping” investments among banks, creating the conditions for indirect financial contagion. These dynamics within banking networks create feedback loops: as banks buy or sell assets to manage their risk, they inadvertently influence the market prices of these assets, further impacting other institutions’ risk levels \cite{Adrian2009, Greenwood2012}, who might then have to buy or sell assets to manage their new risk profile. The key question is whether this process leads to a widespread crisis or stabilizes after initial turbulence.

In current literature, most models assume a homogeneous investment approach, where banks allocate funds evenly across a set number of assets (the so-called $1/N$ rule). However, empirical evidence suggests that banks often adopt more sophisticated strategies, balancing risk and return in line with mean-variance optimization \cite{Elton1997}. Such heterogeneous investment behaviors may affect the overall stability of the network, requiring models that account for varied portfolio weights across institutions.

To investigate the effect of heterogeneous investment behaviors, we build upon the dynamic model developed by Corsi et al. in \cite{CorsiMain} (see \ref{Price Process} for details). This model includes one main feedback mechanism:  portfolio rebalancing. This involves banks adjusting their portfolios to maintain targeted exposure levels as asset prices fluctuate. By simulating these behaviors, the model captures how banks’ portfolio adjustments amplify changes in asset prices, creating feedback loops that can destabilize the financial system \cite{Danielsson2004,Wagner2011}. We extend their model by incorporating heterogeneous investment sizes, enabling institutions to allocate varying amounts of money across different assets.

This added feature has a drastic effect on the stability/instability of the system. Our results indicate that increased heterogeneity correlates with heightened instability, driving the system towards excessive price volatility. By controlling for overall investment levels across institutions, we are able to isolate the effect of heterogeneous investment size, highlighting its role in determining market stability. 

Additionally, when studying the impact of diversification on market stability in a broad range of bank to asset ratios, we identify a \emph{connectivity-driven} behavior. Increasing diversification has two competing effects, each becoming prominent in different connectivity regimes. The first is a stabilizing effect -- prominent at lower connectivity -- which increases the market capitalization of each asset, thus making their prices more resilient to trades. The other is a destabilizing effect - prominent at higher connectivity - which is the result of increased portfolio overlap between banks facilitating the spread of shocks. Hence, the model is most stable at moderate levels of diversification and low levels of heterogeneity. Therefore in order to promote stability, regulators should implement policies that (i) encourage a minimum level of diversification, but discourage over-diversification that would instead lead to more instability, and (ii) discourage institutions from making a small series of very large investments (corresponding to the highest heterogeneity).

It turns out that the stability/instability transition is governed by the average largest eigenvalue of an associated random matrix $\bm{\Phi}$, which dictates the evolution of the dynamical price process. Evaluating this average exactly is quite challenging. We therefore resort to two approximation schemes, as well as direct numerical diagonalization. The two approximate methods differ in their accuracy at different places of the phase diagram. The one proposed in \cite{CorsiMain} tends to underestimate the risk present in the system, while a new approximation we have developed, which leverages the replica method developed in \cite{OurPaper}, tends to more accurately characterize the transition lines, while sometimes erring on the side of caution.

This paper is organized as follows: Section \ref{sec:lit review} presents a brief review of the literature on financial contagion. In Section \ref{sec:Model setup}, we outline the model and introduce investment size heterogeneity into the system. We also demonstrate the connection between financial market stability and the largest eigenvalue of a matrix that encodes the dynamics of the market. Section \ref{sec:results} presents the results, highlighting the impact of investment size heterogeneity on market stability, as well as the impact of diversification. In Section \ref{sec:Methods}, we describe the two methods employed to analyze the average largest eigenvalue of the matrix. Section \ref{Conclusions} is devoted to concluding remarks, while in the Appendices we provide technical details about the model and analytical derivations.

\section{Literature Review}\label{sec:lit review}

Our study intersects with several key areas in the literature on financial stability and systemic risk, each contributing to our understanding of how regulatory, market, and structural factors shape financial vulnerabilities.

First, we build upon a rich body of research examining the effects of capital requirements on the behavior of financial institutions and their potential procyclical impact. Notable works in this domain, such as those by Danielsson et al. \cite{Danielsson2004,Danielsson2009}, have explored the mechanisms by which regulatory capital frameworks might inadvertently exacerbate financial cycles, leading to amplified systemic risks. Adrian and Shin \cite{Adrian2009,Adrian2010,Adrian2014} contribute critical insights into the liquidity channel, examining how bank capital influences asset market dynamics through leverage adjustments. Similarly, Adrian et al. \cite{Adrian2011} and Adrian and Boyarchenko \cite{Adrian2012} have developed theoretical models highlighting the interplay between capital requirements and risk-taking behaviors in financial markets.

Second, we draw extensively from the literature on distressed selling and its effects on market price dynamics, a crucial area that has elucidated how fire-sale mechanisms contribute to volatility and risk widespread contagion. Foundational studies by Shleifer and Vishny \cite{Shleifer1992} and Kyle and Xiong \cite{Kyle2001} laid the groundwork for understanding how forced asset liquidations can precipitate downward price spirals, aggravating financial instability. Building on these findings, Cont and Wagalath \cite{Cont2011,Cont2012} have proposed quantitative models that analyze liquidity shortages and their impact on asset prices. Additional contributions by Thurner et al. \cite{Thurner2012} and Caccioli et al. \cite{Caccioli2015} have further enriched our understanding of endogenous market risks, illustrating how feedback effects from fire sales can increase systemic fragility. Greenwood et al. \cite{Greenwood2012} and Duarte and Eisenbach \cite{Duarte2013} extend these theoretical frameworks to develop systemic risk measures that capture the economy-wide repercussions of fire-sale dynamics.

Third, we consider the effects of diversification and overlapping portfolios on systemic risk. A paradox identified by Wagner \cite{Wagner2011} and later expanded on by Tasca and Battiston \cite{Tasca2011}, Caccioli et al. \cite{Caccioli2012b}, and Lillo and Pirino \cite{Lillo2015} suggests that increased diversification, rather than mitigating systemic risk, can synchronize portfolio responses across institutions, thus elevating aggregate systemic risk. While Wagner \cite{Wagner2011} hypothesizes that such diversification may intensify exposure to common shocks, our study introduces an additional mechanism whereby heterogeneity in investment sizes amplifies systemic risk.

Additionally, our research engages with the literature examining the unintended risks of financial innovation. Studies by Brock et al. \cite{Brock2009} and Caccioli et al. \cite{Caccioli2009} highlight that while innovation can drive financial efficiency, it also introduces novel sources of systemic risk. Haldane and May \cite{Haldane2011} describe how financial products, by creating new linkages and dependencies within the financial system, open up additional channels for contagion that were previously non-existent.

Finally, we contribute to the body of work investigating the dynamics of balance sheet aggregates and credit supply among financial institutions. Foundational studies by Stein \cite{Stein1998} and Bernanke and Gertler \cite{Bernanke1989} have established the critical role of financial intermediaries in transmitting economic shocks. Bernanke et al. \cite{Bernanke1996,Bernanke1999} and Kiyotaki and Moore \cite{Kiyotaki1997} extend this analysis by modeling the feedback loops between asset prices and lending capacity, providing a macroeconomic perspective on how credit cycles impact broader financial stability. Our study builds upon these insights, offering a detailed view of how variations in balance sheet dynamics influence systemic risk across institutions and market conditions.

\color{black}
\section{Model Setup} \label{sec:Model setup}
\subsection{Settings, Notation and Assumptions}
\begin{figure}
    \centering
    \includegraphics[width=0.5\linewidth]{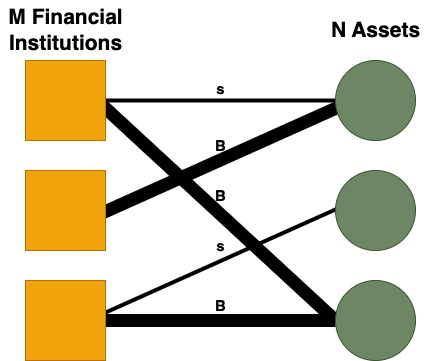}
    \caption{Figure schematically illustrating the bipartite network setup of our model.}
    \label{Schematic Representation}
\end{figure}

\textcolor{black}{We briefly review here the model of Corsi et al. \cite{CorsiMain} highlighting the main point of departure.} Consider a random model of $N$ assets and $M$ financial institutions. We say that each financial institution (labeled $1\leq j \leq M$ ) decides whether or not to invest in an asset (which we label $1\leq i \leq N$ ) with probability equal to $q/\sqrt{NM}$. \textcolor{black}{If the decision to invest is taken, the institution underwrites} $K$ monetary units \textcolor{black}{for the chosen asset}, where $K$ is drawn from a probability \textcolor{black}{density function} $p\left(K\right)$.  The parameter $q$ therefore controls the level of \emph{diversification} in the system, while $p(K)$ encodes its \emph{heterogeneity}. For a single instance of the process, we define the $N\times M$ matrix $\bm{X}$, whose entries $X_{ij}$ represent the investment of the $j$\textsuperscript{th} institution in the $i$\textsuperscript{th} asset. \textcolor{black}{These entries} are random variables (RVs) $X_{ij}=c_{ij}K_{ij}$\textcolor{black}{, where} $c_{ij}\sim \mathrm{Bern}(q/\sqrt{NM})$ indicates whether institution $j$ invested in asset $i$\textcolor{black}{, while} $K_{ij}\sim p(K)$ represents the weight of the investment. This setup is shown schematically in Figure \ref{Schematic Representation}.

Given the above definitions, the entries of the random matrix $\bm{X}$ are drawn from a probability distribution given by

\begin{equation}
    P\left( X_{ij} \right) = \left[\frac{q}{\sqrt{NM}}\delta_{c_{ij},1}+\left(1-\frac{q}{\sqrt{NM}}\right)\delta_{c_{ij},0}\right]p\left(K_{ij}\right).
    \label{eq:X distribution}
\end{equation}

\noindent {The normalization $\sqrt{NM}$ \textcolor{black}{is chosen} so that the probability of investing in an asset depend on the overall ``size'' of the financial market. The intuition is that as the number of financial institutions $N$ increases, \textcolor{black}{one is} less likely to invest in an asset due to the increased competition for that asset from other institutions.} Throughout this paper we restrict ourselves to the case of \textcolor{black}{a binary choice between }light/heavy \textcolor{black}{(Small/Big)} investments, \textcolor{black}{with} the weight distribution \textcolor{black}{taking} the form\footnote{This is the main point of departure from Corsi \cite{CorsiMain}, who instead considered the case in which each institution splits its wealth uniformly among assets it invested in. This is known as the homogeneous $1/N$ rule \cite{Pflug2012,DeMiguel2009}.}

\begin{equation}
    p\left(K\right) = p_B \delta_{K,B} + p_s \delta_{K,s}\ ,\label{eq:defModelpKBs}
\end{equation}

\noindent with 
\begin{equation}\label{eq:constrain_1}
    p_B+p_s=1\ \ .
\end{equation}
Increasing $B$ and $s$ generally raises instability, as it injects more capital into the system. To isolate the impact of investment size heterogeneity, we impose the additional constraint
\begin{equation}\label{eq:constraint_2}
    p_BB+p_ss=1\ .
\end{equation}
The natural requirement that $B\geq1\geq s>0$ follows from the two constraints above. Constraint \eqref{eq:constraint_2} normalizes the amount expected to be invested back to the original homogeneous model \cite{CorsiMain}\footnote{The homogeneous model \cite{CorsiMain} corresponds to Eq. \eqref{eq:defModelpKBs} reducing to $p(K)=\delta_{K,1}$.}, allowing us to examine volatility induced solely by varying investment strategies. Effectively, institutions are expected to invest the same total amount on average but can distribute it differently. 

One may easily generalize our results to capture a richer form of weights distribution.
Furthermore, \textcolor{black}{we will always work in} the sparse regime, meaning that the total number of investments is much smaller than the $N\times M$ possible ones. \textcolor{black}{This natural} assumption implies that the parameter $q$, which fixes the density of non-zero elements in $\bm{X}$, does not scale with either of $\bm{X}$'s dimensions, $N$ and $M$. For all practical purposes, it means that our analysis holds when $q\ll\sqrt{NM}$\footnote{Numerical simulations do not require this condition, and are exact in all settings.}. 

\subsection{Relating Financial Stability with the Largest Eigenvalue} \label{Reltating it to Eigenvalue}

In order to study the stability and properties of \textcolor{black}{the} financial markets, we first have to define \textcolor{black}{the} price dynamics. \textcolor{black}{Following Corsi \cite{CorsiMain},} the price movement has two components: an endogenous component driven by trading at the previous time step called $\bm{e}_{t-1}$, and a random exogenous component called $\bm{\varepsilon}_t$. The \textcolor{black}{vector of asset returns} is given by 
\begin{equation}
{\bm r}_t=\bm{e}_{t-1}+\bm{\varepsilon}_t\ .
\end{equation}
Since the exogenous component is random\textcolor{black}{, it can be written as} $\varepsilon_{i,t}=f_i+\nu_{i,t}$, with the factor $f_i\sim\mathcal{N}(0,\,\sigma_f^{2})$ and idiosyncratic noise $\nu_{i,t}\sim\mathcal{N}(0,\,\sigma_\nu^{2})$ independent. Thus, $\text{Var}(\varepsilon_{i,t})=\sigma_f^2+\sigma_\nu^2$. 
Since this variance is constant in time, we are interested in determining the endogenous return in order to study its volatility. \textcolor{black}{It turns out (see \ref{Price Process} for details)} that the endogenous components \textcolor{black}{${\bm e}_t$ evolve in time as}
\begin{equation}\label{eq:Price Process at start}
    {\bm e}_t =\bm{\Phi}\left(\bm{e}_{t-1}+\bm{\varepsilon}_t\right)\ ,
\end{equation}
\textcolor{black}{where the matrix $\bm\Phi$ reads}
\begin{equation}\label{Phi Master}
    \bm{\Phi}= \frac{(\eta-1)}{\gamma} \alpha^2\bm{W}\bm{W}^T\ ,
\end{equation}
\textcolor{black}{with} $\eta$ the leverage (set by regulatory constraints), $\gamma$ the asset liquidity, $\alpha=\sqrt{\frac{N}{M}}$ the ``structure'' parameter, and $\bm W$ the weight matrix. Both $\eta$ and $\gamma$ are assumed to be the same for all financial institutions and assets, respectively. In this model, the regulatory leverage is given by
\begin{equation}\label{eq:TargetLeverage}
    \eta=\frac{1}{\zeta\sqrt{\sigma_s^2+\frac{\sigma_d^2}{\alpha q}}}\ ,
\end{equation}
where $\zeta$ is a proxy for an institutions risk appetite, $\sigma_s$ is associated with systematic risk, and $\sigma_d$ is the risk that can be mitigated through institutional diversification. The details of the calculations surrounding the regulatory leverage can be found in \ref{Price Process}. An institution's actual leverage is given by $L_{j,t}=\frac{A_{j,t}}{E_{j,t}}$, where $A_{j,t}$ is the amount of asset held by institution $j$ at time $t$, and $E_{j,t}$ is the total amount of equity institution $j$ has at time $t$. It is the difference between the regulatory leverage $\eta$ and the actual leverage $L_{j,t}$ that causes institution $j$ to trade at time $t$. The \textcolor{black}{weight} matrix $\bm{W}$ is of size $N\times M$, and it represents the proportion of asset $i$  in institution $j$'s portfolio. \textcolor{black}{Its} entries are \textcolor{black}{thus} defined as 
\begin{equation}\label{W definition}
    W_{ij}=\frac{X_{ij}}{\sum_{l=1}^NX_{lj}}\ ,
\end{equation} 
\noindent where the $X_{ij}$ follow the probability distribution defined \textcolor{black}{in Eq.}\eqref{eq:X distribution}. We note that the matrix $\bm W$ is column stochastic, \textcolor{black}{with $\sum_{i=1}^N W_{ij}=1$ $ \forall j$}.

\textcolor{black}{Diagonalizing the square symmetric matrix $\bm\Phi$ as $\bm\Phi={\bm P}\bm\Lambda {\bm P}^{-1}$, with $\bm\Lambda$ the diagonal matrix of real eigenvalues $\lambda_i$, and defining $\tilde{{\bm e}}_t={\bm P}^{-1}{\bm e}_t$ and $\tilde{\bm\varepsilon}_t=\bm{P}^{-1}\bm\varepsilon_t$, we can rewrite Eq. \eqref{eq:Price Process at start} for a single asset as}

\begin{equation}\label{equation10}
    \tilde{e}_{t,i}=\lambda_i\tilde{e}_{t-1,i}+\lambda_i\tilde{\varepsilon}_{t,i}\ .
\end{equation}
Recursively iterating equation \eqref{equation10}, we have 
\begin{equation}
    \tilde{e}_{t,i}=\lambda_i^t\tilde{e}_{0,i}+\sum_{k=0}^{t-1}\lambda_i^{t-k}\tilde{\varepsilon}_{k,i}\ .
\end{equation}
\textcolor{black}{Computing the asset price fluctuations, we get
\begin{equation}  \text{Var}\left(\tilde{e}_{t,i}\right)=\sum_{k=0}^{t-1}\lambda_i^{2\left(t-k\right)}\text{Var}\left(\tilde{\epsilon}_{k,i}\right)
=\left(\sigma_f^2+\sigma_{\nu}^2\right)\sum_{j=1}^{t}\lambda_i^{2j}\ .
\end{equation}}

Hence, we see that if even just one of the $\lambda_i>1$, then as $t\rightarrow\infty$ at least one of the fluctuations of the stock return processes will diverge.

\textcolor{black}{The system's typical stability profile is therefore governed by the average \emph{largest} eigenvalue $\mathbb{E}[\lambda_{max}]$ of the random matrix $\bm\Phi$, where $\mathbb{E}[\lambda_{max}]>1$ denotes an unstable market situation characterized by unbounded volatilities. }

The question, therefore, becomes whether the average largest eigenvalue of $\bm\Phi$ can be accurately estimated. One may follow three methods in principle:
\begin{enumerate}[Method \#1:]
\item Numerical generation and diagonalization of random matrices $\bm\Phi$. We refer to this method as numerical diagonalization.
\item Approximate analytical evaluation of the largest eigenvalue of the average matrix $\mathbb{E}[\bm\Phi]$. This strategy allows us to extract some analytical dependence on the parameters of the model, and was originally followed in the paper by Corsi et al. \cite{CorsiMain} for the case of homogeneous investments. We will show below  that in the heterogeneous case, this strategy severely distorts the results, and generally fails to accurately pinpoint the transition lines (see Section \ref{sec:results} for details). We refer to this method as the Corsi approximation.
\item Full analytical evaluation of the average largest eigenvalue of an approximate matrix $\hat{\bm\Phi}$ (suitably defined to be ``close'' to the original matrix $\bm\Phi$). This can be achieved following a recent result in the field of random matrices obtained via the replica method (see \cite{OurPaper} and Section \ref{Complex Math} below). We refer to this method as the Replica approximation.
\end{enumerate}
The most accurate method, full analytical evaluation of the average largest eigenvalue of the matrix $\bm\Phi$, is currently out of reach. In the next section, we provide a summary of our main results for the largest eigenvalue and consequences on the stability-instability transition of the heterogeneous model.

\color{black} 

\section{Results}\label{sec:results}

\begin{figure}[h!]
    \centering
    \begin{tabular}{cc}
        \includegraphics[width=0.5\linewidth]{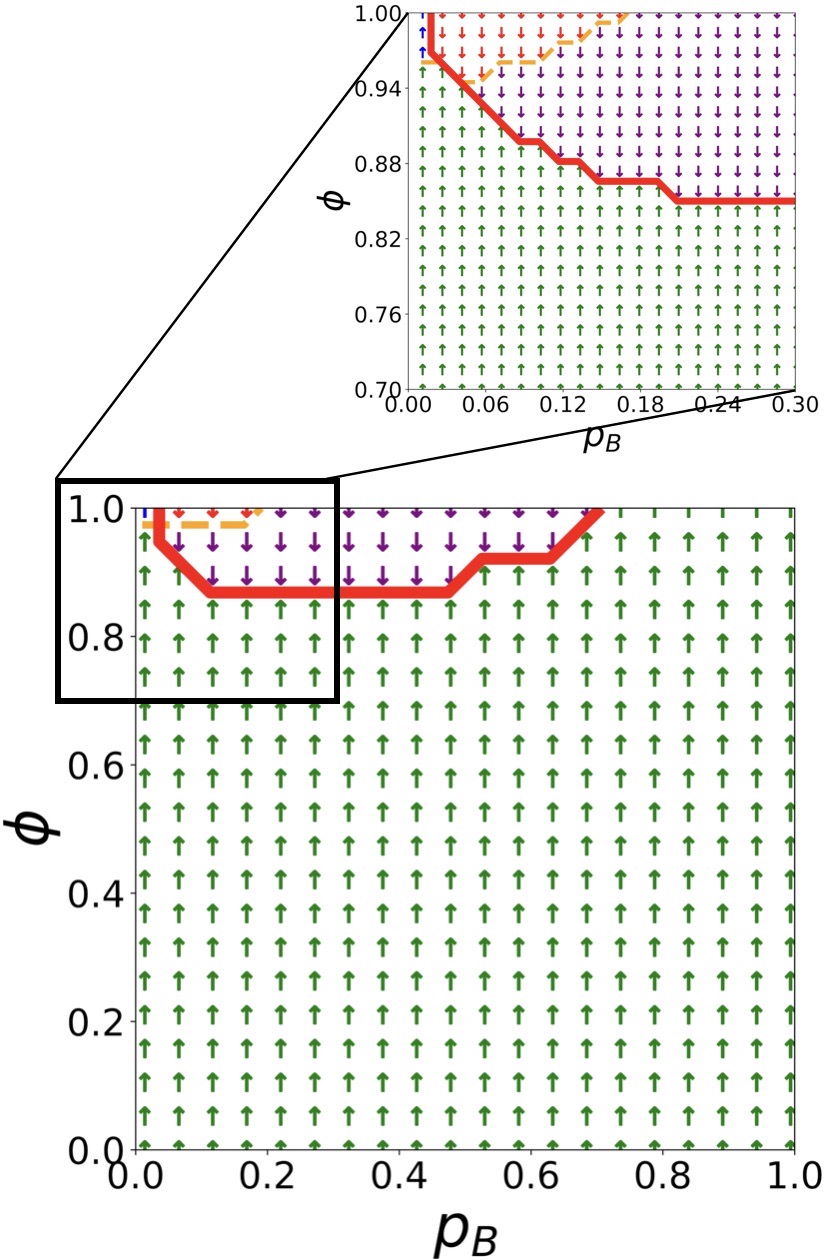}&
        \includegraphics[width=0.5\linewidth]{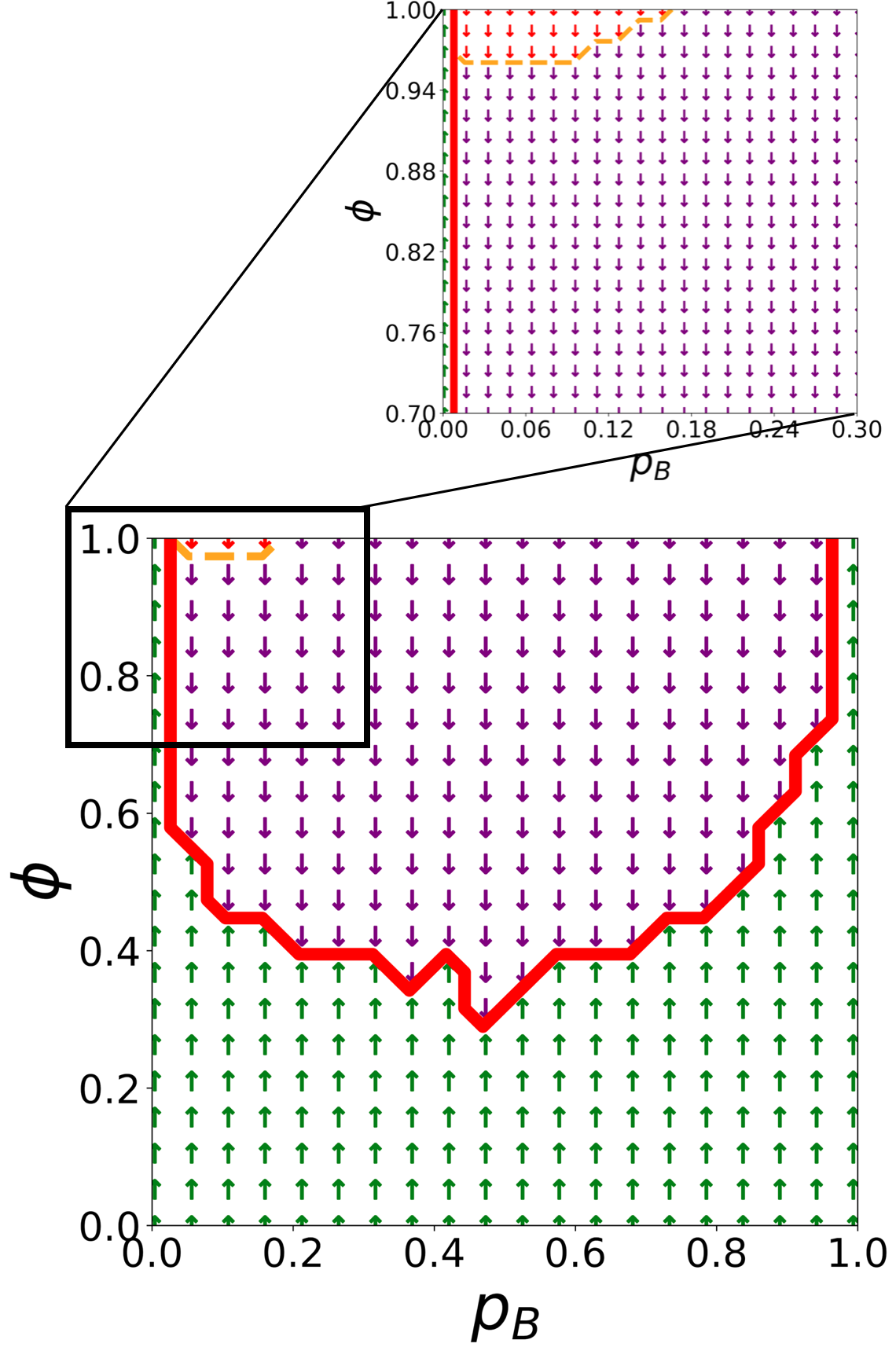} \\
        (a) \(\alpha = \sqrt{\frac{200}{300}}\approx 0.81\) & (b) \(\alpha = \sqrt{\frac{400}{300}}\approx1.15\)
    \end{tabular}
    \caption{Stability analysis of the financial model comparing the Corsi method (method \#2) and direct diagonalization (method \#1) across varying levels of asset structure, measured by the parameter \(\alpha=\sqrt{N/M}\), the amount of heterogeneity $\phi$, and the probability of heavy investment, \(p_B\). The two subfigures represent different values of \(\alpha\) corresponding to (a) low number of assets, and (b) high number of assets. For each configuration, the connectivity parameter is set at \(q = 8\), the risk appetite parameter is set at \(\zeta = 1.85\), the systemic volatility is set at \(\sigma_s^2 = 0.009\), the diversifiable volatility is set at \(\sigma_d^2 = 0.03\), and the liquidity parameter is set at \(\gamma = 50\). The markers represent the stability profile of the market according to Table \ref{arrows} for approximate method \#2.}
    
\label{fig:CORSI_stability_grid}
\end{figure}

\begin{figure}[h!]
    \centering
    \begin{tabular}{cc}
        \includegraphics[width=0.5\linewidth]{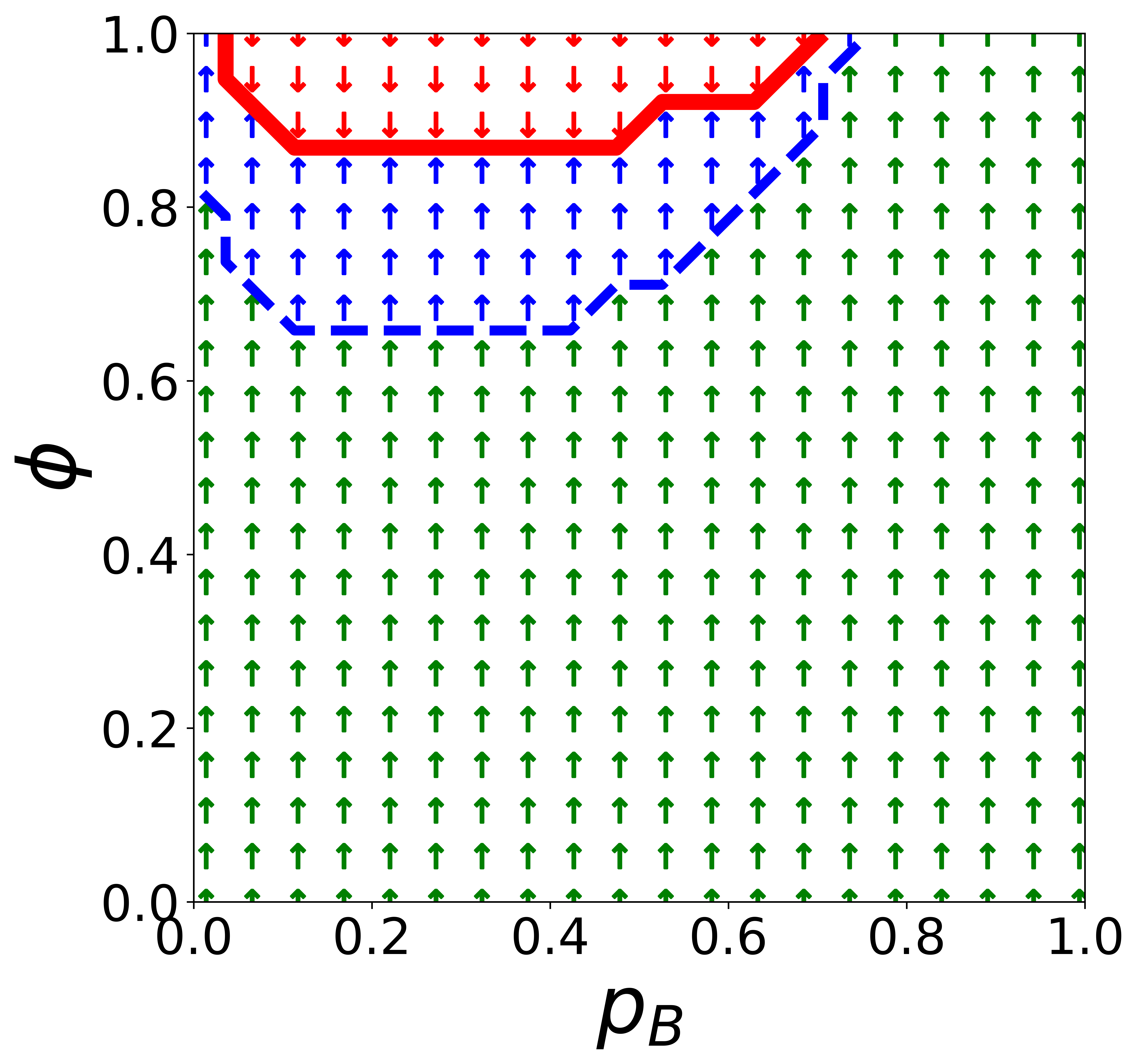} &
        \includegraphics[width=0.5\linewidth]{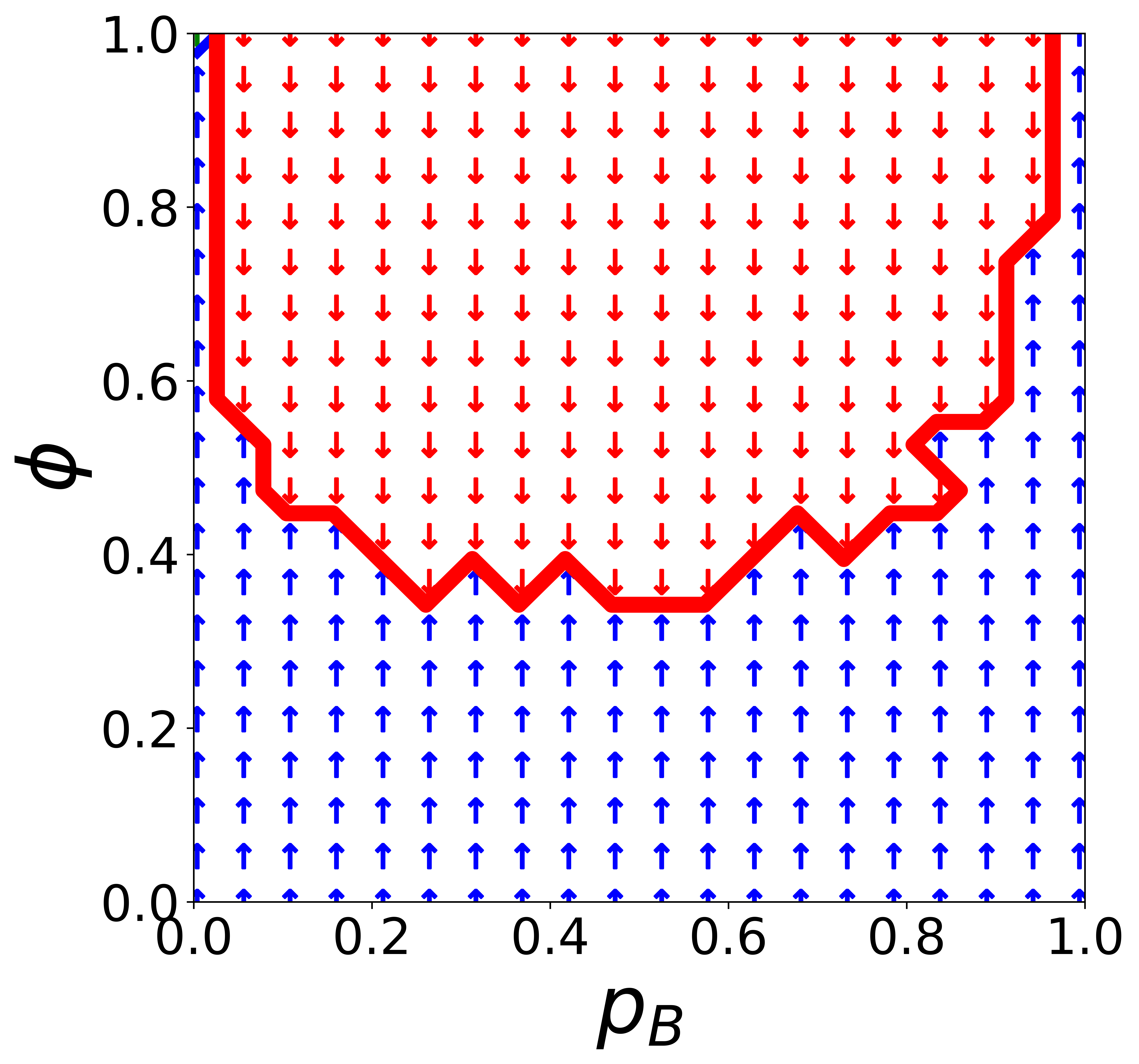} \\
        (a) \(\alpha = \sqrt{\frac{200}{300}}\approx 0.81\) & (b) \(\alpha = \sqrt{\frac{400}{300}}\approx1.15\)
    \end{tabular}
    \caption{Stability analysis of the financial model comparing the replica method (method \#3) and direct diagonalization (method \#1). The parameters are the same as in Figure \ref{fig:CORSI_stability_grid}. The markers represent the stability profile of the market according to Table \ref{arrows} for approximate method \#3.}
    
\label{fig:Replica_stability_grid}
\end{figure}

In the following, we proceed with a numerical analysis focusing on three key aspects of the model. 

First, we analyze the primary question about the influence of investment size heterogeneity on market volatility (see Section \ref{sec:Impact of Heterogeneous Investment Strategies}). This analysis reveals that disregarding the specific manners of institutional investment leads to an underestimation of systemic risk. Notably, there are cases where the system appears stable under a homogeneous investment assumption, yet becomes unstable when heterogeneity in investment sizes is accounted for.

Next, we explore how the model's stability is affected by the connectivity parameter \( q \) that determines the level of diversification (see Section \ref{Impact of Diversification on Market Stability}).  We find that two competing behaviors happen, each at different levels of $q$. For lower values of $q$, increasing diversification promotes stability in the system, whereas for higher values of $q$, it promotes instability.

Finally, we assess the numerical accuracy of method \#2 in comparison to the replica method \#3 (see Section \ref{Comparing the Two Methods}). Our findings indicate that method \#2 tends to underestimate the systemic risk level relative to method \#3.

Our analysis will be focused on two main financial system regimes. The first is when $N<M$, and $\alpha\approx\sqrt{\frac{2}{3}}$. In this regime, the number of assets is smaller than the number of banks in the system. This regime has been studied previously, as it is deemed reflective of many real world financial markets \cite{Duarte2015, Shin2020, Feinstein2023}. The other regime we study is when the number of assets exceeds the number of banks, and thus $N>M$ and $\alpha>1$. This setting is often used in network models of fire sales in order to capture the effect of overlapping portfolios, and is considered a representative model of Globally Systemically Important Banks (G-SIBs) impact on fire sales \cite{Caccioli2012b, Coen2019, Cont2017, Mazarissi2018}. G-SIBs are financial institutions that are large in size and very interconnected, and are deemed to have the potential to cause widespread financial disruption were they to fail \cite{GSIB}.

\subsection{Impact of Heterogeneous Investment Strategies}\label{sec:Impact of Heterogeneous Investment Strategies}
Here, we present the main results of our analysis focusing on the impact of heterogeneous investment strategies in terms of investment sizes. Figure   \ref{fig:CORSI_stability_grid} and Figure \ref{fig:Replica_stability_grid} show the stability profile of the financial market using the two different approximation methods (\#2 and \#3) compared to the actual result from direct numerical diagonalization (method \#1). These figures demonstrate that, irrespective of the method used to evaluate $\mathbb{E}[\lambda_{max}]$, the investment strategies employed by institutions significantly influence system stability and that the method used to evaluate system stability may contribute to a misleading perception of it.

In order to visualize our results, we note that we now have one free parameter $q$ (governing diversification) and four inter-dependent parameters to adjust $(B,s,p_B,p_s)$, such that Eqs. \eqref{eq:constrain_1} and \eqref{eq:constraint_2} are satisfied. 

We can produce illustrative phase diagrams by plotting a normalized indicator of heterogeneity $\phi$ vs. $p_B$, where 
\begin{equation} \label{heterogeneity constraint}
    \phi=1-\frac{s}{B}\ ,
\end{equation}
with $\phi\to 0$ when heterogeneity is small ($B\approx s$) and $\phi\to 1$ when heterogeneity is maximal ($B\gg s$). The model of reference \cite{CorsiMain} is precisely retrieved (i) along the $\phi=0$ line (lower boundary of the frame) due to \eqref{heterogeneity constraint}, (ii) exactly along the $p_B=0$ line (left boundary of the frame), and (iii) along the $p_B=1$ line (right boundary of the frame) due to our constraints \eqref{eq:constrain_1} and \eqref{eq:constraint_2}\footnote{For instance, $p_B=0$ implies $p_s=1$, which in turn forces $s=1$ and thus $p(K)=\delta_{K,1}$ (and similarly for $p_B=1$).}. \begin{comment}, and $p_B\to 0$ (left edge of the frame). \end{comment} Every point on the graph uniquely fixes a set of $(B,s,p_B,p_s)$. 

\begin{figure}[h!]
    \centering
    \begin{tabular}{cc}
        \includegraphics[width=0.5\linewidth]{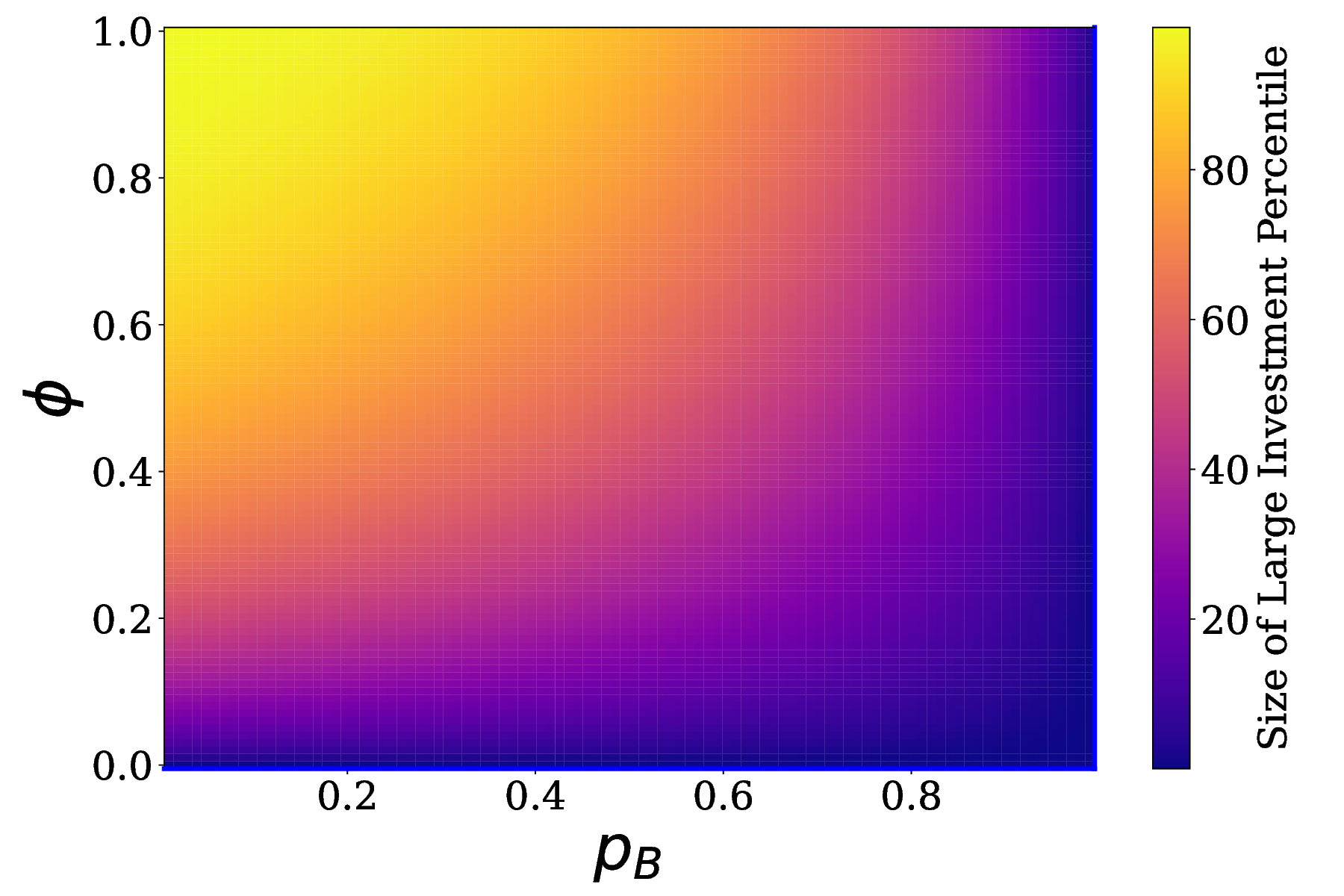} &
        \includegraphics[width=0.5\linewidth]{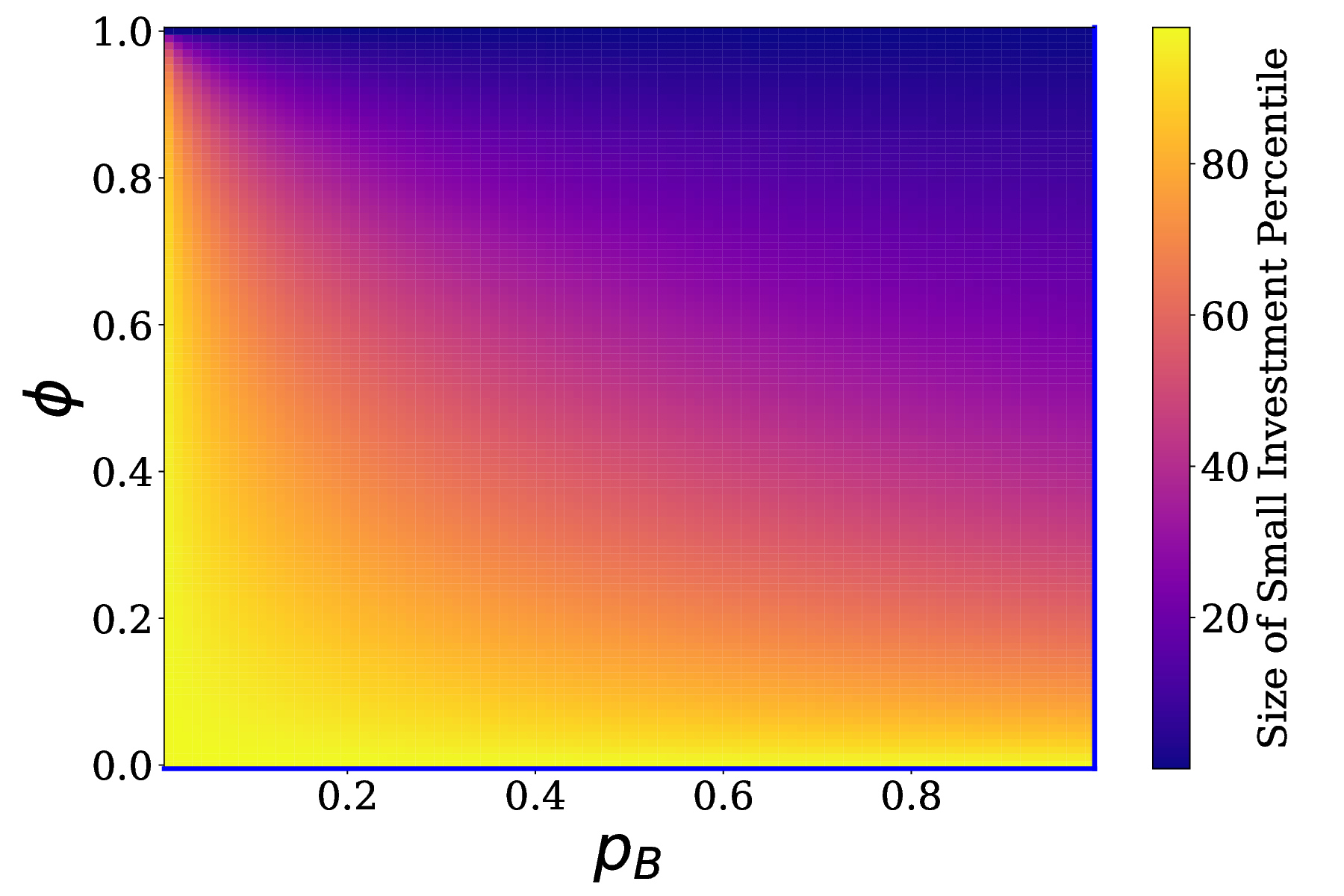} \\
        (a) \(\alpha = \sqrt{\frac{200}{300}}\approx 0.81\) & (b) \(\alpha = \sqrt{\frac{400}{300}}\approx1.15\)
    \end{tabular}
    \caption{Figure showing the evolution of the parameters $B$ (left) and $s$ (right) as a function of the level of heterogeneity $\phi$ and $p_B$.}
    \label{B and s evolution}
\end{figure}

Figure \ref{B and s evolution} illustrates how the parameters \( B \) (size of the large investment, left heatmap) and \( s \) (size of the small investment, right heatmap) evolve as functions of the probability of investing heavily \( p_B \), and the level of heterogeneity \( \phi \).  
The constraint in Equation \eqref{eq:constraint_2} implies that the size of the large investment \( B \) is inversely related to the probability \( p_B \). This explains why the size of $B$ decreases as we move eastwards along the heatmap. Conversely, $B$ increases as we increase the level of heterogeneity and move northwards. An analogous behavior is observed for $s$, except that $s$ shrinks when heterogeneity is increased.

\begin{table}[h!]
    \centering
    \renewcommand{\arraystretch}{1.5} 
    \begin{tabular}{|c|p{10cm}|} \hline
    \textbf{Arrow} & \textbf{Meaning} \\ \hline
    \textcolor{red}{$\downarrow$} & The system is \textbf{unstable}. The approximate method (method \#2 in Figure \ref{fig:CORSI_stability_grid} and method \#3 in Figure \ref{fig:Replica_stability_grid})  used correctly identifies the system as unstable. \\ \hline
    \textcolor{mypurple}{$\downarrow$} &  The system is \textbf{unstable}. The approximate method used incorrectly identifies the system as stable. \\ \hline
    \textcolor{blue}{$\uparrow$} &  The system is \textbf{stable}. The approximate method used incorrectly identifies the system as unstable. \\ \hline 
    \textcolor{mygreen}{$\uparrow$} &  The system is \textbf{stable}. The approximate method used correctly identifies the system as stable. \\ \hline
    \end{tabular}
    \caption{Table summarizing what the colors and arrows in Figures \ref{fig:CORSI_stability_grid} and \ref{fig:Replica_stability_grid} represent.}
    \label{arrows}
\end{table}

We note that both $B$ and $s$ are decreasing functions of $p_B$. Since the constraint in Equation \eqref{eq:constraint_2} must be met, a large $p_B$ situation means that banks are forced to lower the size of their large investments, which however happens more frequently. Thus, the overall amount invested stays the same on average.

Turning to Figures \ref{fig:CORSI_stability_grid} and \ref{fig:Replica_stability_grid}, we see a clear transition between the region where $\mathbb{E}[\lambda_{max}]>1$ and the system is unstable, and the region where $\mathbb{E}[\lambda_{max}]<1$ and the system is stable. However different methods estimate different transition lines.  The transition line between actual instability and stability (as measured by method \#1) is the red line, whereas the orange dashed line represents the transition line between stability and instability according to the two approximate methods (method \#2 in Figure \ref{fig:CORSI_stability_grid}, and method \#3 in Figure \ref{fig:Replica_stability_grid}). We see that instability predominantly appears in the upper left corner, representing high heterogeneity, while stability prevails in more homogeneous settings at the graph's bottom. This suggests that disregarding heterogeneity and assuming financial institutions invest homogeneously would lead to underestimating the risk present in the financial system, and thus that introducing heterogeneity is an important aspect to better study the stability of financial markets.

For a certain level of relatively high heterogeneity $\phi$, increasing the probability $p_B$ of  large investment has a stabilizing effect. This phenomenon arises from the inverse relation between $p_B$ and $B$, as discussed above. This explains why the rightmost edge of the diagram is stable for higher levels of heterogeneity than in the middle of the graph, because in this direction the size of the large investment shrinks.

We also notice that in the leftmost region of the graph, the system is stable for higher values of $\phi$ than at medium values of $p_B$. This is due to the shrinking possibility of investing heavily, as well as the model resembling the homogeneous setting in the extreme $p_B\to 0$ limit. However, the system can still be unstable for very low (but nonzero) values of $p_B$, and very high levels of heterogeneity (uppermost left corner), as this region implies a high value of $B$, as discussed above.

The destabilizing impact of investment heterogeneity can be understood through the dual roles of small and large investments within the interconnected financial network. In this framework, smaller investments serve as conduits that propagate shocks across institutions, facilitating the transmission of financial stress. Larger investments then amplify these shocks, by creating large market fluctuations as these heavy positions are subject to more substantial buying and selling pressures when institutions adjust their portfolios.

This effect is intrinsic to our model, which extends previous frameworks by accounting for shock propagation via overlapping investments of varying sizes. While a shock to an asset held in small quantities may impose limited direct financial strain on an institution’s balance sheet, it nonetheless triggers portfolio rebalancing to maintain regulatory leverages. In this rebalancing process, institutions sell shares of both their heavily and lightly invested assets. Consequently, assets with larger investment sizes experience outsized sell-offs, exerting stronger downward pressure on prices than would occur under a homogeneous investment distribution.

Thus, heterogeneity increases the sensitivity of the price process to institutional actions, as price declines in heavily weighted assets become more pronounced. This underscores that investment heterogeneity, by magnifying both direct and indirect channels of financial contagion, renders the market more vulnerable to instability.

\subsection{Impact of Diversification on Market Stability}\label{Impact of Diversification on Market Stability}

\begin{figure}[h!]
    \centering
    \begin{subfigure}{0.5\textwidth}
        \centering
        \includegraphics[width=\linewidth]{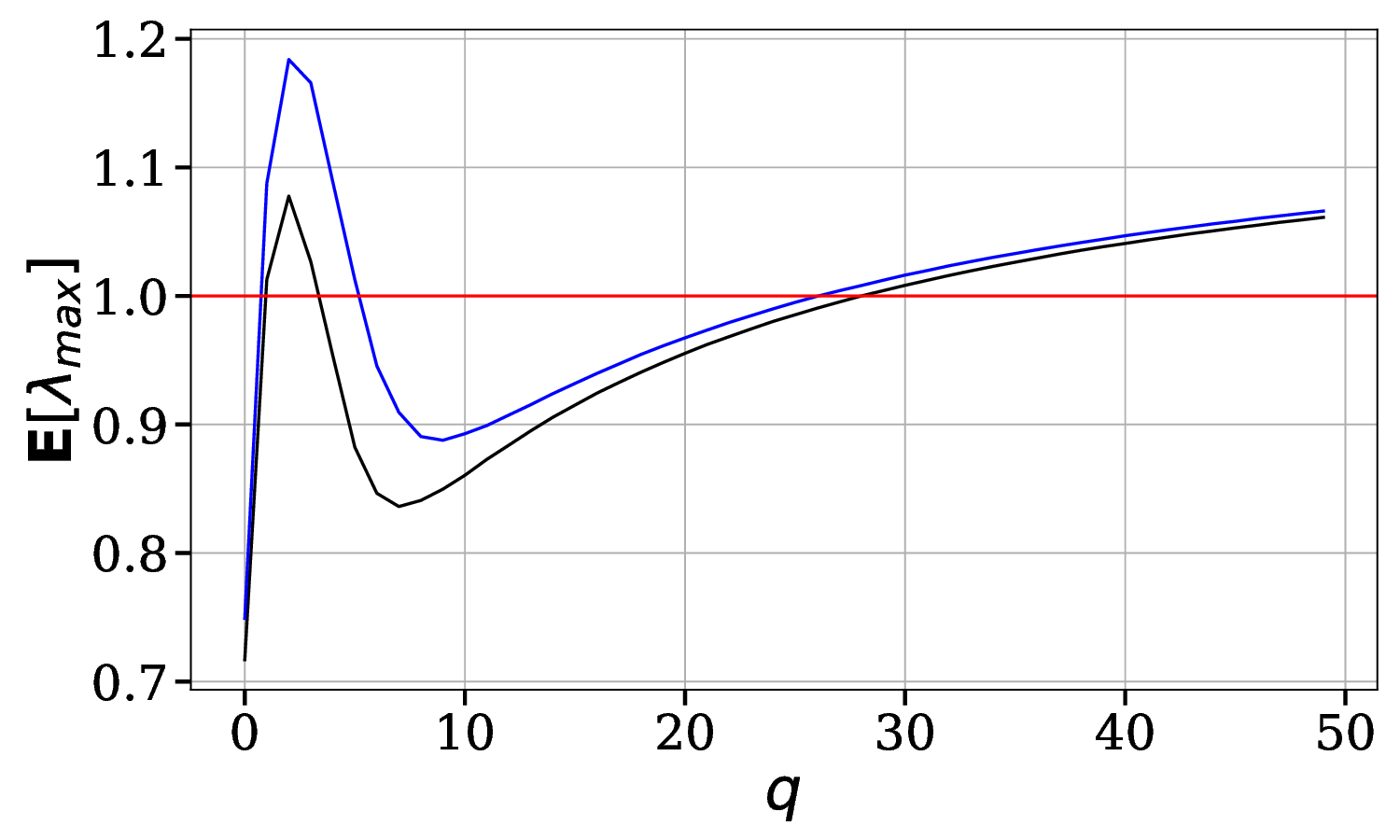}
    \end{subfigure}%
    \hfill
    \begin{subfigure}{0.5\textwidth}
        \centering
        \includegraphics[width=\linewidth]{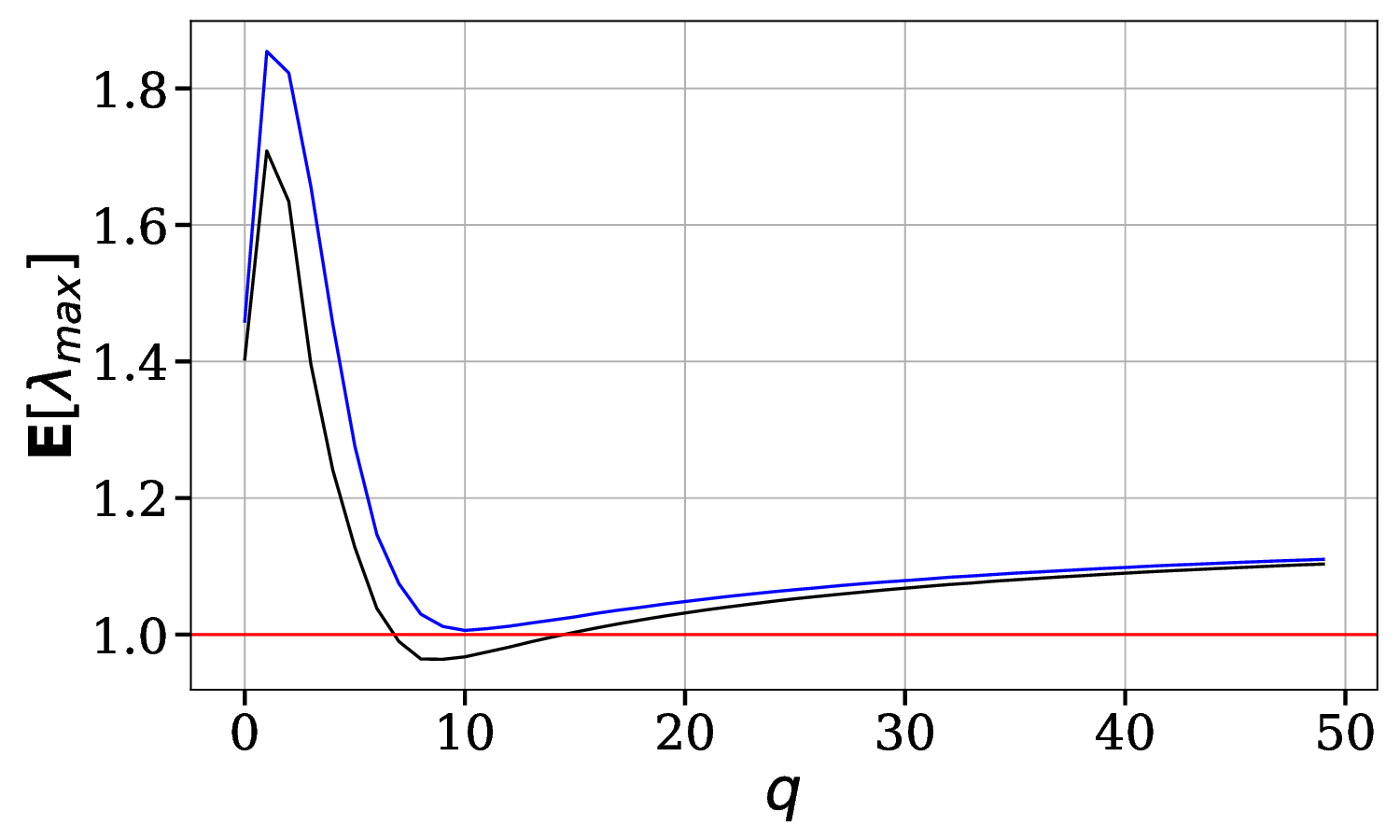}
    \end{subfigure}
    \caption{Figures showing the evolution of the average largest eigenvalue of the matrix $\bm\Phi$ as a function of the connectivity parameter $q$. The blue line represents our heterogeneous case with the parameters set at $B=3$, $s=0.3$, $p_B=\frac{7}{27}\simeq 0.26$, and $p_s=\frac{20}{27}\simeq 0.74$. The black line represents the homogeneous case with $B=s=1$. Other parameters are the same as in Figure \ref{fig:CORSI_stability_grid}. The red line represents the instability threshold of $\mathbb{E}\left[\lambda_{max}\right]=1$. Left: the lower number of assets regime ($\alpha=\sqrt{200/300}$). Right: the higher number of assets regime ($\alpha=\sqrt{400/300}$).}
    \label{fig:both_q_summary}
\end{figure}

In Figure \ref{fig:both_q_summary}, we analyze the effect of institutional diversification $q$ on the average largest eigenvalue of the matrix $\bm\Phi$ and the volatility of the financial system. 
Consistently with previous literature that showed a non-monotonic behavior of systemic stability with respect to diversification \cite{Gai2010,Caccioli2012b}, the figure shows the existence of a peak in the expected value of $\mathbb{E}\left[\lambda_{max}\right]$, which can be intuitively understood in terms of two competing effects: on one hand, the increase in diversification makes individual institutions more stable. On the other hand, it also provides more paths for the propagation of instabilities.

After this peak, the system becomes consistently more unstable as $q$ increases. This is due to the following reason: In the model, as $q$ increases, the average number of assets held by each institution increases according to $q\sqrt{\frac{N}{M}}$, thus increasing diversification and portfolio overlap between institutions. At the same time, leverage also increases, according to equation \eqref{eq:TargetLeverage}, as banks can increase leverage because diversification benefits offset the associated risks. This increase in leverage explains the increasing trend observed in the figure.

We study the behavior of our model in two distinct regimes. The first regime, with $\alpha=\sqrt{\frac{200}{300}}$,  corresponds to a scenario in which the number of assets is smaller than the number of institutions investing in them.  Figure \ref{fig:both_q_summary} (left) shows how the average largest eigenvalue of the matrix $\bm\Phi$ evolves as a function of the connectivity parameter $q$ in this setting using direct diagonalization (method \#1). Figure \ref{fig:both_q_summary} (right) shows the regime in which the number of assets is larger than the number of banks. Here, $\alpha=\sqrt{\frac{400}{300}}$, which models the G-SIBs setting.

We note that the average largest eigenvalue in the G-SIBs regime typically reaches higher values than the regime with lower $\alpha$. From the perspective of individual assets, each asset has
\begin{equation}
    \mathbb{E}\left[\sum_{j=1}^MX_{ij}\right]=\frac{q}{\alpha}
\end{equation}
institutions investing in it on average. When $\alpha>1$, each asset is typically held by a very small number of institutions at low values of  $q$. As a result, the market share of each institution is disproportionately large, regardless of whether the institution has a heavy or light investment in the asset, as it is one of the few shareholders. Consequently, when an institution buys or sells an asset in this regime, the price impact is substantial, since such transactions represent a significant proportion of the overall asset volume. In this setting, the price fluctuations are therefore significantly larger, as each trade has an outsized effect on the price. This explains the higher average largest eigenvalue (and thus greater instability) in the $N>M$ regime.

We also note that in both cases, instability peaks around low values of $q$. This is also related to the relative market share each institution holds in each available asset. As $q$ is increased, the amount of volatility initially shrinks rapidly. This is because as $q$ is increased, each asset receives more investors, which in turn increases the market capitalization of each asset, allowing it to absorb trades without such extreme price fluctuations (see the price process given by Eq. \eqref{exogenous process} in \ref{Price Process}). Hence, initially diversification is beneficial to market stability as it allows for market depth to be created. However, after a certain point, the drawbacks of diversification as found in \cite{CorsiMain} resurface, as increased overlap in portfolios allows for shocks to spread through the system. Hence, after a minimum at around $q\approx10$, the average largest eigenvalue becomes an increasing function of $q$, and the system is progressively pushed towards higher levels of instability.

\subsection{Comparing the Two Methods}\label{Comparing the Two Methods}

We now compare the results obtained using method \#2, described in Section~\ref{Corsi}, with those obtained from method \#3, introduced in Section~\ref{Complex Math}. Our analysis reveals that the methods in Section~\ref{Corsi} tend to underestimate the level of risk present in the system when compared to the more accurate methods presented in Section~\ref{Complex Math}, which tend to accurately predict when the market is going to be unstable, but also tend to slightly overestimate the risk present in the system. Figures \ref{fig:CORSI_stability_grid} and \ref{fig:Replica_stability_grid} allow us to directly compare the two methods.

Figure \ref{fig:CORSI_stability_grid} shows that method \#2 (Corsi method) severely underestimates the amount of risk present in both settings, as shown by the large regions of purple down arrows. These arrows represent that direct numerical diagonalization predicts an unstable market, but Equation \eqref{lambda max master} predicts a stable market instead. Indeed, method \#2 only manages to accurately predict that the system is unstable in the most extreme setting of heterogeneity (upper left corner of the graph, and the inset). Hence, in most settings, using this methodology would lead to an incorrect determination of the stability of the financial markets. Additionally, as discussed in Section \ref{Impact of Diversification on Market Stability}, increasing the structure parameter $\alpha$ should lead to more instability, as each institution's market share in each asset increases, and thus each trade has a larger impact on price movements. This is not the case in Figure \ref{fig:CORSI_stability_grid}, as the region predicted to be unstable is actually \emph{smaller} in the inset of panel (b), corresponding to higher $\alpha$, than it is in panel (a) which corresponds to lower $\alpha$. Thus not only does this method underestimate the risk of heterogeneity, it also fails to pick up the additional instability resulting from higher values of $\alpha$.

Figure \ref{fig:Replica_stability_grid} shows us that method \#3 (Replica method) is more accurate than method \#2, and importantly, accurately predicts when the system is unstable. This is shown by the lack of purple down arrows, and the presence of only red downwards arrows. However, it is not a perfect approximation, and overestimates the amount of risk in the system, shown by the blue upwards arrows. In these regions, the replica method approximation deems the system unstable, while in reality it is stable. From a policy point of view however, this mis-classification would be less significant as it can be interpreted as an abundance of caution. Additionally, this methodology accurately picks up on the additional risk due to increasing values of $\alpha$.

While both methods agree that investment size heterogeneity increases \(\mathbb{E}[\lambda_{max}]\) and consequently market variance, the improved estimate of \(\mathbb{E}[\lambda_{max}]\) reveals that the impact of increasing the disparity in investment sizes is more pronounced than previously indicated by method \#2. Additionally, from a policy point of view, ``missing'' by overestimating the risk present is preferable to underestimating it. Using method \#3 will never result in a false stable diagnosis, and thus abiding by the stability/instability regions of method \#3 ensures stability.

\section{Methods} \label{sec:Methods}

In this section, we briefly review method \#2, which involves calculating the largest eigenvalue of the average matrix $\mathbb{E}[\bm\Phi]$, and \#3, which involves analytically solving for the average largest eigenvalue of an approximate matrix $\tilde{\bm{\Phi}}$ described below.

\subsection{Method \#2 (\cite{CorsiMain})} \label{Corsi}

In this section, we will study the heterogeneous model using the same approximate methodology as in \cite{CorsiMain}: estimating the average largest eigenvalue of the random process via the largest eigenvalue of the average process. 

We recall that the original price process is determined by a random matrix $\bm\Phi$ given by (see Eq. \eqref{Phi Master})
\begin{equation}\label{Phi Estimating Master}
    \bm\Phi=\frac{(\eta-1)}{\gamma} \frac{N}{M}\bm{W}\bm{W}^T\ .
\end{equation}

We start by defining the average process 
\begin{equation}\label{expected process}
\tilde{\bm e}_t=\mathbb{E}[\bm\Phi](\tilde{\bm e}_{t-1}+\bm \varepsilon_t)\ ,
\end{equation}
and studying its dynamics.

We find that the diagonal and off-diagonal terms of the average matrix are given by
\begin{align}
    \mathbb{E}\left[\Phi_{ii}\right]&=\frac{\eta-1}{\gamma}\frac{b}{q\sqrt{MN}}\ ,\label{Phi diagonal terms}
    \\
    \mathbb{E}\left[\Phi_{ij}\right]&=\frac{\eta-1}{\gamma}\frac{1}{M\left(\sqrt{MN}b-q\left(1-N\right)\right)}\ ,\label{Phi off diagonal terms}
\end{align}
where from now on we write  $b=B^2p_B+s^2p_s$. We find that the largest eigenvalue $\tilde\lambda_{max}$ of $\mathbb{E}\left[\bm\Phi\right]$ can be found in closed form as
\begin{equation}\label{lambda max master}
\tilde\lambda_{max}=\frac{\eta-1}{\gamma}\left(\frac{b}{q\alpha}+\frac{q}{\frac{b}{\alpha}+q}\right) \ ,
\end{equation} 
in the large $N,M$ limit. For the details of the calculations, see \ref{Corsi Methodology Calculations}.

\color{black}

\begin{figure}[h!]
\centering
        \includegraphics[width=\linewidth]{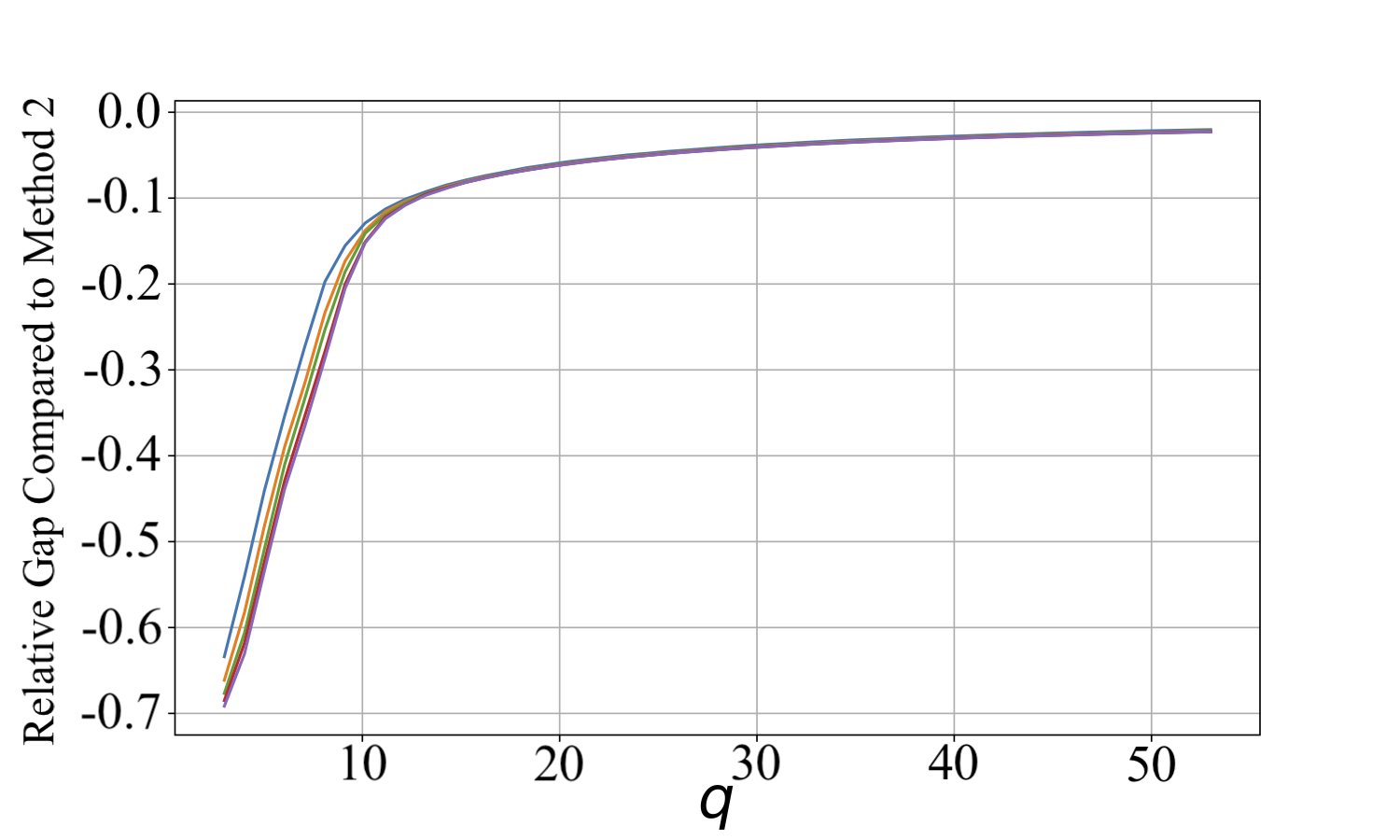}

    \caption{Graph showing the relative gap between $\tilde\lambda_{max}$ as predicted by Equation \eqref{lambda max master} and the numerical results from direct diagonalization, for different scales of matrices. The parameters are the same as Figure \ref{fig:both_q_summary}.  The structure parameter is set at $\alpha=\sqrt{\frac{400}{300}}$, and thus in this instance the matrix is size $400d\times 300d$, and the colors represent different values of $d$.}\label{fig:Corsi Gap}
\end{figure}

Figure \ref{fig:Corsi Gap} shows the relative difference between the approximated $\tilde\lambda_{max}$ of Equation \eqref{lambda max master} and the value of $\mathbb{E}[\lambda_{max}]$ obtained from averaging over $10^4$ realizations of the random matrix $\bm\Phi$. It is evident that while the approximate value $\tilde\lambda_{max}$ approaches the exact value as the diversification parameter $q$ increases, notable differences persist. These differences are particularly pronounced in the sparse investment regime (lower $q$ values), where the estimated $\tilde\lambda_{max}$ deviates sharply from the numerical results. At low $q$, this approximation initially underestimates the true eigenvalue by nearly $70\%$. This early-stage underestimation suggests that, in practical terms, reliance on this formula alone might lead to a substantial under-assessment of systemic risk in a financial context -- potentially compromising stability analyses. 

The chart suggests that while method \#2 serves as a useful starting point for analytical purposes, the compounded approximations contribute to a formula that is not sufficiently precise, as observed when comparing the phase diagrams of Figure \ref{fig:Replica_stability_grid} and Figure \ref{fig:CORSI_stability_grid}. A similar problem is present in the homogeneous case discussed in \cite{CorsiMain}, however the conclusions drawn there about stability/instability are unaffected by the lack of precision in the determination of the largest eigenvalue. This is because, while the precise value of the largest eigenvalue is determined with low accuracy, it is never misclassified as being $>1$ or $<1$ in the ``wrong'' region of the phase diagram. However, the consequences of this approximation being quite crude are much more severe in the heterogeneous case, as shown by the large area of incorrectly identified regions in Figure \ref{fig:CORSI_stability_grid}. 

In the next subsection, we show that a better analytical control over the largest eigenvalue can be achieved resorting to a recent replica calculation \cite{OurPaper}.

\color{black}

\subsection{Method \#3 (Replica)}\label{Complex Math}

Motivated by the differences between analytical results obtained through method \#2 and numerical results, in the following section we will present an analytical approach based on the replica method \cite{Nagao2006,Kuehn2008, Susca2021} to study the average largest eigenvalue, $\mathbb{E}[\lambda_{max}] $, of a square $N\times N$ symmetric matrix $\bm{J}$. This problem can be formulated in terms of the Courant-Fisher optimization problem

\begin{equation}
    \lambda_{max} = \frac{1}{N}\max_{\bm{v}\in \mathbb{R}^N,\ |\bm{v}|^2=N} \left(\bm{v},\bm{Jv}\right)\ ,
    \label{eq:Lambda1 Opt}
\end{equation}
where $(\cdot,\cdot)$ stands for the standard dot product.

For large matrix size $N$, it is convenient to recast the optimization problem by introducing a fictitious temperature $\beta$ and the canonical partition function
\begin{equation}
     Z(\beta)=\int d\bm{v} \exp\left(\frac{\beta}{2}\left(\bm{v},\bm{Jv}\right)\right)\delta\left(|\bm{v}|^2-N\right)\ ,
\end{equation}
where the integral runs over components of the $N$-dimensional normalized vector $\bm v$.

In the zero temperature limit, $\beta\rightarrow \infty$, the Gibbs measure $P\left(\bm{v}\right) \propto \exp\left(\frac{\beta}{2}\left(\bm{v},\bm{Jv}\right)\right)$ concentrates around the ground state (i.e., the only contribution to the integral comes from the top eigenvector of $\bm{J}$) and the free energy reads

\begin{equation}
    F\left(\beta\rightarrow \infty\right) = \lim_{\beta\rightarrow\infty} \frac{1}{\beta}\ln Z(\beta) = \lim_{\beta\rightarrow\infty} \frac{1}{\beta}\ln ~e^{\beta \max (\bm v,\bm J\bm v)/2} = \frac{N\lambda_{max}}{2}\ ,
\end{equation}
where we assumed that the largest eigenvalue is not degenerate. Averaging over different realizations of $\bm{J}$, we obtain

\begin{equation}
    \mathbb{E}[\lambda_{max}]  = \lim_{\beta\rightarrow\infty} \frac{2}{\beta N} \mathbb{E}[ \ln Z(\beta)] \ . 
\end{equation}

\noindent The average is then computed using the replica trick,

\begin{equation}
    \mathbb{E}[\lambda_{max}] = \lim_{\beta\rightarrow\infty} \frac{2}{\beta N}  \lim_{n\rightarrow 0} \frac{1}{n}\ln\mathbb{E}[Z(\beta)^n ]\ ,
    \label{eq:Lambda_Replica}
\end{equation}

\noindent where $n$ is initially treated as an integer, and then analytically continued to real values around $n=0$. The limit $N\to\infty$ is also understood to be taken before the replica limit.

Applying this method directly to the matrix $\bm J = \bm W \bm W^T$ is difficult, due to the correlated structure of the entries of the matrix $\bm W$ (see Eq. \eqref{W definition})\footnote{This evaluation would correspond to directly evaluating the average largest eigenvalue of $\bm\Phi$.}. However, we can approximate  \( \bm{W} \) by \( c\bm{X} \), where \( c \) is a constant to be determined. This approximation (method \#3) reformulates the process in terms of the matrix \( \bm{X}\bm{X}^T \), where the entries of \( \bm{X} \) (defined in Equation \eqref{eq:X distribution}) are independent and sparse, allowing us to directly apply the replica results of \cite{OurPaper} to the present case.

The constant $c$ can be determined by imposing that the first moment of the entries of our approximate matrix  equals the first moment of the entries of the actual matrix $\bm W$, leading to
\begin{equation}\label{constant}
    c=\frac{\mathbb{E}[W_{ij}]}{\mathbb{E}[X_{ij}]}\ .
\end{equation}
We find in \ref{estimate calculations} that our approximated price process is thus given by
\begin{equation}\label{approximated Phi}
    \bm\Phi\sim \frac{(\eta-1)}{\gamma} \frac{\left(1-e^{-\alpha q}\right)^2}{q^2}\bm{X}\bm{X}^T=\kappa\bm{X}\bm{X}^T\ .
\end{equation}

The average largest eigenvalue in the replica setting is determined via the solution of recursive distributional equations (integral equations for auxiliary probability density functions), which can be efficiently obtained using a population dynamics algorithm (see \cite{OurPaper} for details).

\begin{figure}[h!]
    \centering
    \begin{subfigure}{0.5\textwidth}
        \centering
        \includegraphics[width=\linewidth]{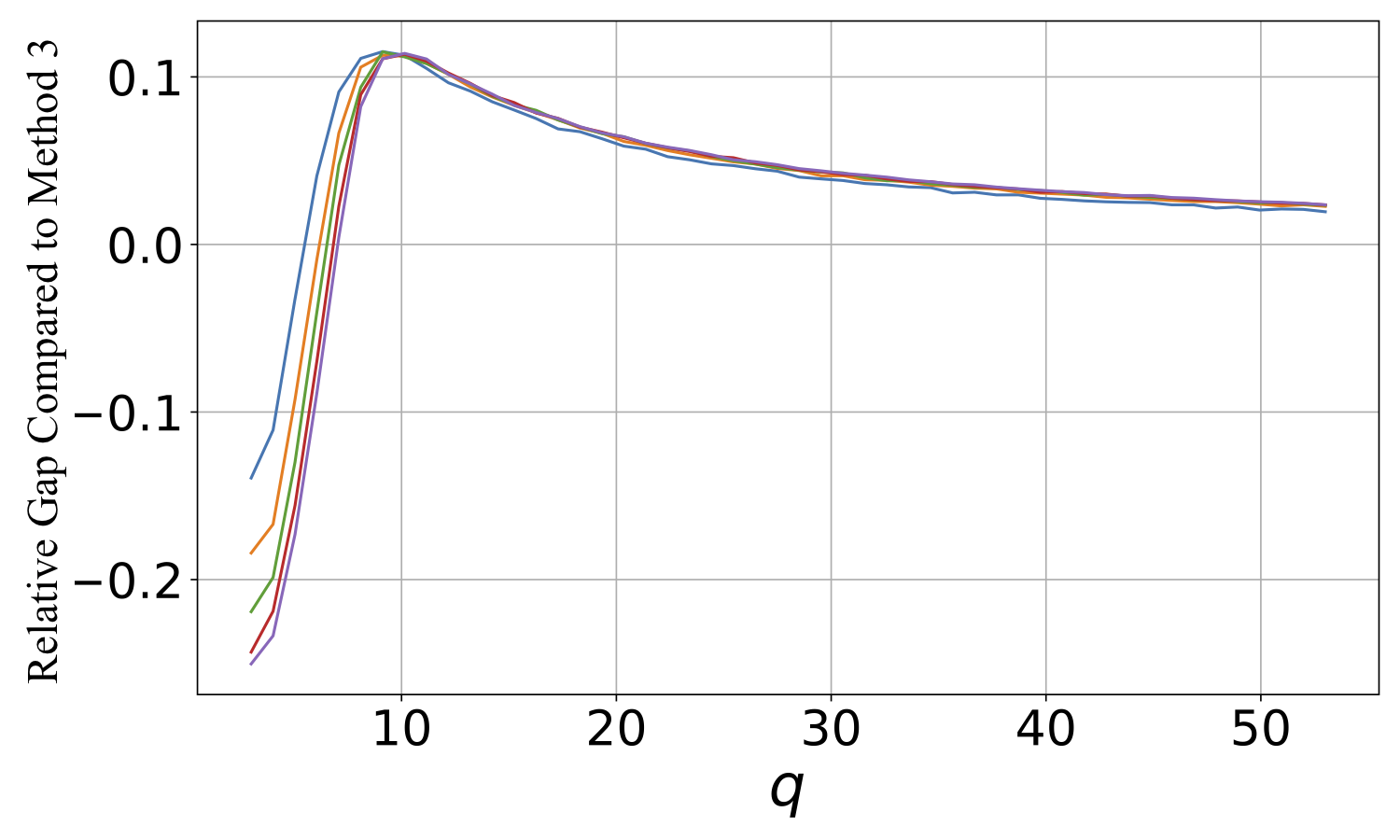}
        \label{fig:gap_200_300}
    \end{subfigure}%
    \hfill
    \begin{subfigure}{0.5\textwidth}
        \centering
        \includegraphics[width=\linewidth]{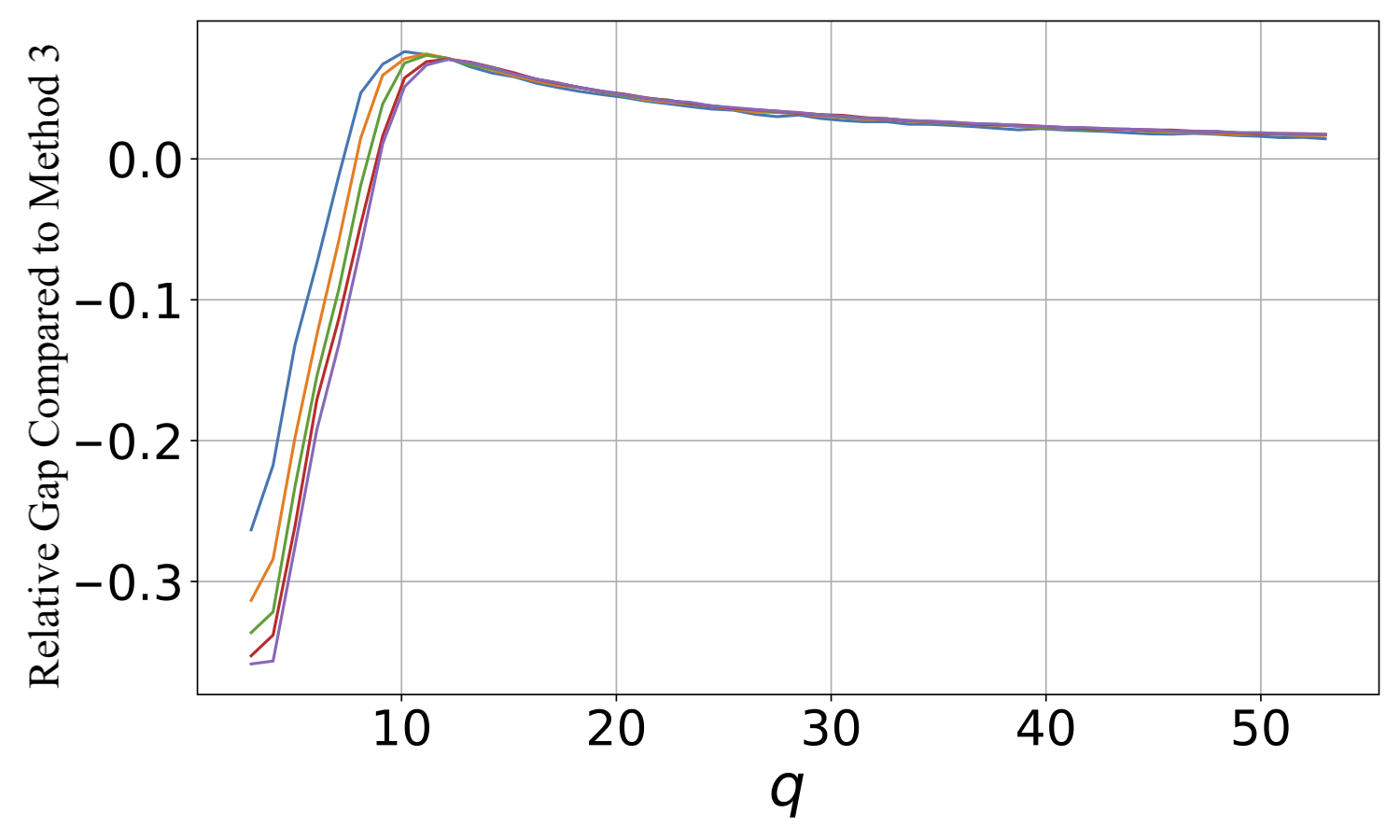}
        \label{fig:gap_500_300}
    \end{subfigure}
    \caption{Graphs showing the change in relative gap as a function of $q$ between the 
    average largest eigenvalue of the approximate matrix $\bm\Phi$ in Eq. \eqref{approximated Phi} computed using the replica method and population dynamics (see \cite{OurPaper}) and the average largest eigenvalue of the exact process in Eq. \eqref{Phi Master} evaluated by numerical diagonalization of $10^4$ samples of $\bm X$.  The Figure on the left sets $\alpha=\sqrt{\frac{200}{300}}$  and the Figure on the right sets $\alpha=\sqrt{\frac{400}{300}}$. The other parameters are as in Fig. \ref{fig:both_q_summary}, and the colors represent the scale parameter as in Fig. \ref{fig:Corsi Gap}}. \label{fig:both_images}
\end{figure}

We have tested the accuracy of this estimation of $\bm\Phi$ numerically, and the results are shown in Figure \ref{fig:both_images}. These figures show the evolution of the relative gap between the largest eigenvalue of the estimated matrix $\bm\Phi$ from Equation \eqref{approximated Phi} and the exact eigenvalue from Equation \eqref{eq:Price Process at start} for increasing values of the connectivity parameter $q$ and for different values of the ``scale'' parameter $d$. 
We see that the scale parameter $d$ has almost no bearing on the accuracy of the approximation. Additionally, as the connectivity increases the approximation becomes more accurate. Similar observations were found for other values of the model parameters. Hence, in order to apply the results of \cite{OurPaper}, we can select a value of $q$ for which our approximation is under 5\% off, and then scale up the matrix in order to preserve the sparsity condition that is required by the replica method calculations.

\section{Conclusion} \label{Conclusions}
In this paper, we analyzed the stability of financial investment networks, where financial institutions hold overlapping portfolios of assets, and investigated the impact of heterogeneous investments and portfolio diversification on financial stability. Building on the random matrix dynamical framework proposed by \cite{CorsiMain}, we introduce heterogeneity in the investment behavior, allowing banks to choose between light ($s$) and heavy ($B$) investments with probabilities $p_s$ and $p_B$ respectively. We also studied the effect of connectivity/diversification (parametrized by $q$) on the systemic stability. Following \cite{CorsiMain}, we connected the stability/instability transition, where the assets' volatility starts growing unbounded, to the average largest eigenvalue of the matrix $\bm\Phi$ governing the evolution of the endogenous component of the returns. 

We found that the heterogeneity of investment sizes plays a crucial role in undermining the stability of financial markets. Smaller investments facilitate the propagation of shocks between institutions, while larger investments amplify price fluctuations, as institutions tend to sell larger quantities of these assets during periods of market stress. Consequently, neglecting the heterogeneity of investment sizes leads to a significant underestimation of the systemic risk and potential instability present in the financial ecosystem.

Consistently with previous literature, we also found that increasing diversification can lead to higher levels of volatility in the market. However, when the connectivity parameter $q$ is low, increasing diversification initially reduces systemic risk by creating market depth. This is due to the large relative market shares owned by financial institutions of each asset for low $q$, and thus institutions' trades have a much larger impact on the price. Hence, increasing diversification does initially lower market-wide risk by allowing for assets' market capitalization to grow and thus for the price impact of trades to be smaller. However, after reaching a minimum around $q\approx10$, the negatives of increased portfolio overlap due to diversification reappear, and we return to the monotonically increasing function of $q$ found in previous literature.

Moreover, we showed that the analytical determination of the average largest eigenvalue of $\bm\Phi$ requires some approximations, which however do not lead to consistent estimates. In particular, the strategy proposed in \cite{CorsiMain}, which estimates $\mathbb{E}[\lambda_{max}]$ by calculating the largest eigenvalue of the average process $\mathbb{E}[\bm{\Phi}]$, yields a severe underestimation of risk in the heterogeneous setting. We contrasted it with results from a replica calculation and with numerical diagonalization of large randomly generated matrices to produce the phase diagrams in Section \ref{sec:results}. The new methodology accurately predicted regions where the system was unstable, but tended to overestimate the instability present. However, this is less of a severe ``miss'', as using this methodology would never lead to a false sense of stability.

The model presented here could be extended in several directions, using different investment distributions and connectivities, analyzing the fluctuations of the largest eigenvalue (not just its average), and introducing direct interactions between financial institutions on top of the indirect connections through the commonly shared assets. 

\section*{Acknowledgments}
P.V. acknowledges support from UKRI FLF Scheme (No. MR/X023028/1).

\newpage

\appendix

\section{Derivation of Eqs. \eqref{eq:Price Process at start} and \eqref{Phi Master}}\label{Price Process}
In order to study how trading affects asset prices, one models the endogenous component of price movement $e_{i,t}$ through a simple linear price impact model following \cite{Caccioli2015, Bouchaud1998, Delpini2019}. This price impact process takes into account three components. The first is asset liquidity; the second is the amount being traded; and the third is the market capitalization of the asset, which measures the overall quantity of asset in the market. The price impact process is thus given by
\begin{equation}\label{exogenous process}e_{i,t}=\frac{1}{\gamma}\frac{d_{i,t}}{\chi_{i,t}}\ ,
\end{equation}
where $\chi_{i,t}$ is the market capitalization of asset $i$ at time $t$,  $d_{i,t}$ the amount being traded at time $t$, and $\gamma$ is a constant representing liquidity. This function tries to capture the relative amount of an asset being traded. If a large fraction of the asset is sold $\left(\frac{d_{i,t}}{\chi_{i,t}}\sim 1\right)$, then the price will move substantially. Additionally, if the asset is very illiquid $(\gamma \ \text{small})$, then it will also move substantially. Conversely, if the asset is liquid $(\gamma \ \text{large})$, or has a large market capitalization $\chi_{i,t}$, then the price will move less, as there is more depth in the market to absorb impacts from trades.

We begin by determining the total amount of asset $i$ traded at time $t$. To do so, we introduce the concept of \emph{leverage} of institution $j$ at time $t$, given by:
\begin{equation}\label{leverage}
    L_{j,t}=\frac{A_{j,t}}{E_{j,t}}\ ,
\end{equation}
where $E_{j,t}$ is the total amount of equity institution $j$ has at time $t$, usually defined as assets minus liabilities \cite{Adrian2010,Bardoscia2015, Aymanns2014}, and $A_{j,t}$ is the amount of asset held by bank $j$ at time $t$. Financial institutions are subject to regulations that impose caps on their leverage, such as Basel III. While there are other regulatory constraints institutions must meet, leverage caps are intended to be the ``backstop'' of financial regulation, since they do not consider the risk profile of a firm's holdings, but merely the size. The goal of leverage caps is to bound the default probability of institutions in times of financial stress. Put simply, they force financial institutions to have some capital buffer in order to absorb losses \cite{Basel3}. Hence, given a regulatory leverage $\eta_{j,t}$, we assume institutions maximize their leverage $(L_{j,t}=\eta_{j,t})$ in order to increase potential profits. Hence, an institution's desired amount of asset is given by
\begin{equation}
    A^*_{j,t}=\eta_{j,t} E_{j,t} \ .
\end{equation}
This often differs from the actual amount of asset an institution holds at time $t$, which we have denoted by $A_{j,t}$. This is due to price changes of the individual assets an institution holds. Thus, having to re-balance their portfolio in order to meet their regulatory leverage is what drives the trading dynamics. Hence, the amount of asset that bank $j$ must trade at time $t$ is given by:
\begin{equation}
    D_{j,t} = A_{j,t}^*-A_{j,t}=\eta_{j,t}E_{j,t}-A_{j,t}\ . \label{DeltaA Intermediary}
\end{equation}
Since institutions are considered ``inactive'' between time steps, and we are not modeling liabilities, an institution's equity changes through the profit/loss of its portfolio at time $t-1$. Similarly, its amount of asset $A_{j,t}$ changes depending on the returns of its portfolio at $t-1$. Recalling that $W_{ij}$ represents the proportion of asset $i$ in institution $j$'s portfolio, we can see that the institutions' portfolio returns can be written as
\begin{equation}\label{eq:portfolio_return_vector}
    r_{j,t}^p=\sum_k^NW_{kj}r_{k,t}\ ,
\end{equation}
where $r_{k,t}$ is the return of asset $k$ at time $t$. Hence,
\begin{align}\label{evolution of asset and equity}
    E_{j,t}=E_{j,t-1}+r_{j,t}^pA_{j,t-1}^*, && A_{j,t}=A_{j,t-1}^*+r_{j,t}^pA_{j,t-1}^*\ .
\end{align}
Inserting \eqref{evolution of asset and equity} into \eqref{DeltaA Intermediary}, and assuming that the regulatory leverage stays constant in time (i.e. $\eta_{j,t}=\eta_j$) we see that

\begin{equation}
     D_{j,t}=A^*_{j,t-1}+\eta_{j}r^p_{j,t}A^*_{j,t-1}-A^*_{j,t-1}-r^p_{j,t}A^*_{j,t-1}=r_{j,t}^pA_{j,t-1}^*(\eta_{j} - 1)\ ,\label{amount traded}
\end{equation}
which is the total amount that each institution $j$ must trade at time $t$ to meet its regulatory leverage.

Institutions trade according to the portfolio weights $W_{ij}$, by trading $W_{ij}D_{j,t}$ amount of asset $i$. We note that this means the amount of asset held by institution $j$ of asset $i$ after trading is given by
\begin{equation}
    X_{ij}'=X_{ij}+W_{ij} D_{j,t} = X_{ij}\left(1+\frac{D_{j,t}}{\sum_rX_{rj}}\right)\ .
\end{equation}
Thus the portfolio weight of asset $i$ after trading reads
\begin{equation}
    W_{ij}'=\frac{X_{ij}'}{\sum_kX_{kj}'}=\frac{X_{ij}\left(1+\frac{D_{j,t}}{\sum_rX_{rj}}\right)}{\sum_{k}X_{kj}\left(1+\frac{D_{j,t}}{\sum_rX_{rj}}\right)}=W_{ij}\ .
\end{equation}
This implies that trading in this manner keeps the portfolio weights constant in time.

Multiplying Equation \eqref{amount traded} by the respective portfolio weights, we see that the amount of asset $i$ traded at time $t$ by institution $j$ is given by
\begin{equation}\label{eq:individual_demand}
    d_{i,j,t}=W_{ij}D_{j,t}=W_{ij}A_{j,t-1}^*(\eta_{j} - 1)r_{j,t}^p\ .
\end{equation}
One further assumes that all institutions have the same regulatory leverage, i.e. $\eta_{j}=\eta \ \forall j$. Hence the total amount of demand in asset $i$ is Equation \eqref{eq:individual_demand} summed over $j$, and thus we can write 
\begin{equation}\label{stock demand}
    \bm{d}_t=(\eta-1)\bm{W}_t \bm{Q}_{t-1}\bm{W}^T\bm{r}_t\ ,
\end{equation}
with $\bm{Q}_{t-1}=\text{diag}\left(A^*_{j,t-1} \right)_{j=1}^M$, $\bm r_t\in\mathbb{R}^N$ the vector of asset returns, and $\bm d_t\in\mathbb{R}^N$ the vector of asset trade volumes.

We now estimate the market capitalization of asset $i$, given by $\chi_{i,t}$. An asset's market capitalization is determined by two factors: the average number of institutions investing in it (given by $\mathcal{N}_i$), and the average amount invested in it by each institution (given by $\bar{I}_{i,t-1}$) at the end of the previous time step. Hence, our estimated market capitalization is given by
\begin{equation}\label{estimated market cap early}
    \chi_{i,t}=\mathcal{N}_{i,t} \cdot \bar{I}_{i,t-1}\ .
\end{equation}
Starting with the number of institutions investing in an asset, on average, each asset has
\begin{equation}\label{number of institutions}
\mathcal{N}_{i,t}=\mathbb{E}\left[\sum_{j=1}^M\delta_{X_{ij,1}}\right]=M\mathbb{E}\left[\delta_{X_{ij,1}}\right]
\end{equation}
institutions investing in it, which is constant in time.  We recall from Equation \eqref{eq:X distribution}, that $\mathbb{E}\left[\delta_{X_{ij,1}}\right]=\frac{q}{\sqrt{MN}}$, and thus working in the large $N$ and $M$ regime with scale parameter fixed $\alpha=\sqrt{\frac{N}{M}}$, we have
\begin{equation}\label{number of investors}
    \mathcal{N}_{i,t}=\mathcal{N}_i=\frac{q}{\alpha}\ .
\end{equation}
We now note that on average each institution is supposed to invest its wealth amongst
\begin{equation}
\mathbb{E}\left[\sum_{i=1}^NX_{ij}\right]=q\alpha 
\end{equation}
number of assets. Each asset $i$ is therefore expected to have
\begin{equation}
    I_{i,j,t-1}=\frac{A^*_{j,t-1}}{q\alpha} \ ,
\end{equation}
monetary units invested in it from bank $j$ at the end of time $t-1$. We use $A_{j,t-1}^*$ as this is the total amount of asset held by bank $j$ at the end of trading at $t-1$, in order to meet the regulatory leverage requirement. Further, we assume that there is an average market-wide institution size given by 
\begin{equation}\label{average institution size}
    \bar{A}_{t-1}^*=\frac{1}{M}\sum_{j=1}^MA_{j,t-1}^*\ .
\end{equation}
Since institutions invest following the same rules, they should, on average, invest the same amount.  Hence we can approximate
\begin{equation}\label{average equal j}
    A_{j,t-1}^*\sim\bar{A}_{t-1}^*\ ,
\end{equation}
implying that each institution holds roughly an equal amount of asset. We can therefore estimate the amount invested in asset $i$ by each bank as
\begin{equation}\label{amount invested in}
    \bar{I}_{i,t-1}= \frac{\bar{A}^*_{t-1}}{q\alpha}\ .
\end{equation}
Inserting \eqref{amount invested in} and \eqref{number of investors} into \eqref{estimated market cap early}, we see that we can estimate the asset market capitalization at time $t$ as
\begin{equation}\label{market cap}
\chi_{i,t}=\frac{1}{\alpha^2}\bar{A}_{t-1}^*\ .
\end{equation}

Hence inserting \eqref{market cap} and \eqref{stock demand} into Equation \eqref{exogenous process}, we see that our price impact process is given by
\begin{equation}
    \bm{e}_t=\frac{\left(\eta-1\right)}{\gamma}\alpha^2\bm{W}\frac{\bm{Q}_{t-1}}{\bar{A^*}_{t-1}}\bm{W}^T\left(\bm{e}_{t-1}+\bm\varepsilon_t\right)\ .
\end{equation}
Recalling $\bm{Q}_{t-1}=\text{diag}\left(A^*_{j,t-1} \right)_{j=1}^M$, and applying the approximation of Equation \eqref{average equal j}, we see that $\frac{\bm{Q}_{t-1}}{\bar{A^*}_{t-1}}\sim \mathds{1}$, and we obtain our expression from Equation \eqref{eq:Price Process at start}:
\begin{equation}\label{eq:Phi_With_W}
     \bm{e}_t=\frac{(\eta-1)}{\gamma} \alpha^2\bm{W}\bm{W}^T\left(\bm{e}_{t-1}+\bm\varepsilon_t\right)=\bm\Phi\left(\bm{e}_{t-1}+\bm\varepsilon_t\right)\ .
\end{equation}
This is valid for any given regulatory leverage. We determine our regulatory leverage by following the work of \cite{CorsiMain, Adrian2010, Adrian2014, Aramonte2023, Aramonte2021}, and assuming that institutions operate under a ``Value-at-Risk'' (VaR) constraint on how leveraged they can be. Mathematically, the Value-at-Risk (VaR) constraint is expressed as
\begin{equation}
    \zeta \sigma_p A \leq E\ ,
\end{equation}
where $\zeta$ is a scaling constant that reflects the institution's risk appetite, and $\sigma_p$ denotes the overall volatility of its portfolio. The intuition behind this constraint is that fluctuations in the portfolio's value should not result in losses that exceed the institution's equity, as this would lead to insolvency. To ensure this, the level of assets $A$ must remain constrained such that the maximum expected adverse market impact, represented by $\zeta \sigma_p$, does not surpass the available equity. 
We assume that institutions saturate this inequality in order to maximize the amount of assets they can invest in, and thus maximize potential returns. Hence the VaR constraint reads
\begin{equation}
    \zeta\sigma_pA=E \ .
\end{equation}
This implies that our regulatory leverage is given by
\begin{equation}
    \eta=\frac{A}{E}=\frac{1}{\zeta\sigma_p} \ , 
\end{equation}
where the overall portfolio risk is given by two components: one that is systematic and immitigable given by $\sigma_s^2$, and the other, given by $\sigma_d^2$, that is reduced by the number of assets invested in, $\alpha q$, through portfolio diversification \cite{Mokkelbost1971, BenHorim1980, Markowitz1950}. Hence
\begin{equation}
    \sigma_p^2=\sigma_s^2+\frac{\sigma_d^2}{\alpha q}\ .
\end{equation}

\section{Method \#2 Derivation}\label{Corsi Methodology Calculations}

In order to arrive at Equations \eqref{Phi diagonal terms} and \eqref{Phi off diagonal terms}, we adapt the methodology in \cite{CorsiMain} to our setting, and estimate the averages and variance of functions of random variables through a Taylor expansions around the mean.  Namely, given two random variables $A$ and $B$, with means $\mathbb{E}[A]=\theta_a$ and $\mathbb{E}[B]=\theta_b$, we can expand a function $f(A,B)$ around $\bm\theta=(\theta_a,\theta_b)$, allowing us to approximate the average of this function to the first order
\begin{align}
    \mathbb{E}[f(A,B)]&\approx\mathbb{E}[f(\bm\theta)+f'_a(\bm\theta)(A-\theta_a)+f'_b(\bm\theta)(B-\theta_b)] \nonumber
    \\
    &=\mathbb{E}[f(\bm\theta)]+f'_a(\bm\theta)\mathbb{E}[A-\theta_a]+f'_b(\bm\theta)\mathbb{E}[B-\theta_b] =f(\bm\theta) \ . \label{expectency taylor expansion}
\end{align}
We can now approximate:
\begin{equation}
   \text{Var}(f(A,B))= \mathbb{E}\left[\left[f(A,B)-\mathbb{E}\left[f(A,B)\right]\right]^2\right]\approx \mathbb{E}\left[\left[f(A,B)-f\left(\bm\theta\right)\right]^2\right] \ .
\end{equation}
We can again Taylor expand $f(A,B)$ around $\bm\theta$ to give:
\begin{align}
    \text{Var}(f(A,B))&\simeq\ \mathbb{E}\left[\left[f(\bm\theta)+f'_a(\bm\theta)(A-\theta_a)+f'_b(\bm\theta)(B-\theta_b)-f(\bm\theta)\right]^2\right] \nonumber
    \\
    &=\mathbb{E}\left[{f'_a}^2(\bm\theta)(A-\theta_a)^2+2f'_a(\bm\theta)f'_b(\bm\theta)(A-\theta_a)(B-\theta_b)+{f'_b}^2(\bm\theta)(B-\theta_b)^2\right] \nonumber
    \\
    &={f'_a}^2(\bm\theta)\text{Var}(A)+2f'_a(\bm\theta)f'_b(\bm\theta)\text{Cov}(A,B)+{f'_b}^2(\bm\theta)\text{Var}(B)\ .\label{variance taylor approx}
\end{align}
Specifically, if $f(A,B)=\frac{A}{B}$, then $f_{a}'=\frac{1}{B}$, and $f'_b=\frac{-A}{B^2}$. Hence in this setting:
\begin{align}
    \text{Var}\left(\frac{A}{B}\right)&\approx\frac{\text{Var}(A)}{\theta_b^2}-2\frac{\theta_a}{\theta_b^3}\text{Cov}(A,B)+\frac{\theta_a^2}{\theta_b^4}\text{Var}(B)
    \\
    &=\frac{\theta_a^2}{\theta_b^2}\left[\frac{\text{Var}(A)}{\theta_a^2}-2\frac{\text{Cov}(A,B)}{\theta_a\theta_b}+\frac{\text{Var}(B)}{\theta_b^2}\right] \ .\label{taylor ratio variance}
\end{align}
Equations \eqref{taylor ratio variance} and \eqref{expectency taylor expansion} are the two approximations we will use in order to obtain our average matrix.

We now note from Equation \eqref{Phi Estimating Master} that:
\begin{align}
    \Phi_{ii}&=\frac{(\eta-1)}{\gamma}\frac{N}{M}\sum_{k=1}^MW_{ik}^2 \ ,\label{Phi Diaganola 4}
    \\
    \Phi_{ij}&=\frac{(\eta-1)}{\gamma} \frac{N}{M}\sum_{k=1}^MW_{ik}W_{jk}\ . \label{Phi off diagonala terms}
\end{align}
In order to arrive at the results of Equations \eqref{Phi diagonal terms} and \eqref{Phi off diagonal terms} we have to find both $\mathbb{E}\left[W_{ij}^2\right]$ and $\mathbb{E}\left[W_{ik}W_{jk}\right]$.  For simplicity we write $n_j=\sum_{r}X_{rj}$ to represent the amount of monetary units institution $j$ invests in total. 

Starting with the cross terms $\mathbb{E}\left[W_{ik}W_{jk}\right]$, we note that:
\begin{equation}
    \mathbb{E}[W_{ik}W_{jk}]=\mathbb{E}\left[\frac{X_{ik}X_{jk}}{n_k^2}\right] \ .
\end{equation}
Setting $A=X_{ik}X_{jk}$, $B=n_k$ and $f(A,B)=\frac{A}{B}$, we can apply approximation \eqref{expectency taylor expansion} to get:
\begin{align}
    \mathbb{E}[W_{ik}W_{jk}]&\approx\frac{\mathbb{E}[X_{ik} X_{jk}]}{\mathbb{E}[n_k^2]}=\frac{\mathbb{E}[X_{ik}]\mathbb{E}[X_{jk}]}{\mathbb{E}[n_k^2]} \ ,\label{inter1}
\end{align}
where we have used the independence of $X_{ik} \ \text{and} \ X_{jk}$.  Recalling Equation \eqref{eq:X distribution}, we see that $\mathbb{E}(X_{ik})=\frac{q}{\sqrt{NM}}$. To compute the average of $n_k^2$, we first note from \eqref{eq:X distribution} that:
\begin{equation}
\text{Var}(X_{ij})=\frac{q}{\sqrt{NM}}b-\frac{q^2}{NM}\ ,
\end{equation}
where we recall $b=p_BB^2+p_ss^2$. Again using the independence of $X_{ij}$, we also have:
\begin{equation}
    \text{Var}(n_k)=N\text{Var}(X_{ik})=q\sqrt{\frac{N}{M}}b-\frac{q^2}{M}\ .\label{Variance sum}
\end{equation}
We thus have:
\begin{align}
    \mathbb{E}[n_k^2]&=\text{Var}(n_k)+\mathbb{E}[n_k]^2=q\sqrt{\frac{N}{M}}b+\frac{q^2}{M}\left(N-1\right) .\label{sum second moment}
\end{align}
Inserting Equation \eqref{sum second moment} into \eqref{inter1} gives us the following approximation for the average of the off diagonal terms:
\begin{equation}
    \mathbb{E}[W_{ik}W_{jk}]\approx\frac{\frac{q^2}{NM}}{q\sqrt{\frac{N}{M}}b+\frac{q^2}{M}\left(N-1\right)}\ . \label{Off diagonal intermediate}
\end{equation}

Coming back to Equation \eqref{Phi Diaganola 4}, we note that:
\begin{equation}\label{variance inter 1}
    \mathbb{E}[W_{ij}^2]=\text{Var}(W_{ij})+\mathbb{E}[W_{ij}]^2\ .
\end{equation}
We use \eqref{expectency taylor expansion} to approximate:
\begin{equation}\label{variance inter 2}
    \mathbb{E}[W_{ij}]^2\approx\left(\frac{\mathbb{E}[X_{ij}]}{\mathbb{E}[n_j]}\right)^2=\frac{1}{N^2} \ .
\end{equation}
For $\text{Var}(W_{ij})$, we use our approximation from Equation \eqref{taylor ratio variance}, with $A=X_{ij}$, and $B=n_j$ to see that
\begin{equation}\label{variance approx new}
\text{Var}(W_{ij})\approx \frac{\mathbb{E}[X_{ij}]^2}{\mathbb{E}[n_j]^2}\left(\frac{\text{Var}(X_{ij})}{\mathbb{E}[X_{ij}]^2}-2\frac{\text{Cov}(X_{ij},n_j)}{\mathbb{E}[X_{ij}]\mathbb{E}[n_j]}+\frac{\text{Var}(n_j)}{\mathbb{E}[n_j]^2}\right)\ .
\end{equation}
Since each bank decides to invest in each asset independently, $\text{Cov}\left(X_{ij},X_{rj}\right)=0$ for $i\neq r$. Thus we have that
\begin{equation}
\text{Cov}(X_{ij},n_j)=\text{Cov}\left(X_{ij},\sum_{r}X_{rj}\right)=\sum_r\text{Cov}(X_{ij},X_{rj})=\text{Var}(X_{ij})\ .
\end{equation}
After inserting the formulas for the variances and averages of $X_{ij}$ and $n_j$, we find that after simplifications
\begin{equation}\label{simplified variance new}
    \text{Var}(W_{ij})\approx\frac{\sqrt{M}b}{qN\sqrt{N}}-\frac{1}{N^2}-\frac{\sqrt{M}b}{qN^2\sqrt{N}}-\frac{2\sqrt{M}}{N^2\sqrt{N}}-\frac{1}{N^3}\ ,
\end{equation}
which can be further simplified keeping only terms up to $\mathcal{O}(\frac{1}{N^2})$ as
\begin{equation}\label{variance inter3}
    \text{Var}\left(W_{ij}\right)\approx\sqrt{\frac{M}{N}}\frac{b}{qN}-\frac{1}{N^2} \ .
\end{equation}
Inserting Equations \eqref{variance inter 2} and \eqref{variance inter3} into \eqref{variance inter 1}, we can approximate the second moment of $W_{ij}$ as
\begin{equation}\label{W squared}
    \mathbb{E}[W_{ij}^2]\approx\sqrt{\frac{M}{N}}\frac{b}{qN}\ .
\end{equation}
Thus, we have the following expectations for the entries of $\bm{W}$
\begin{align}\label{W estimates}
    \mathbb{E}[W_{ij}^2]&=\sqrt{\frac{M}{N}}\frac{b}{qN}\ ,& \mathbb{E}[W_{ik}W_{jk}]=\frac{q}{N\left(\sqrt{MN}b+q\left(N-1\right)\right)}\ .
\end{align}
Recalling Equations \eqref{Phi Diaganola 4} and \eqref{Phi off diagonala terms}, inserting \eqref{W estimates} gives Equations \eqref{Phi diagonal terms} and \eqref{Phi off diagonal terms}.  Therefore our average matrix is given by
\begin{equation}
\mathbb{E}[\bm\Phi]=\begin{bmatrix} 
   gd & gd_o & \dots & gd_o \\
   gd_o & gd & \dots & gd_o \\
    \vdots & & \ddots & \vdots\\ 
   gd_o & \dots & & gd
    \end{bmatrix},
\end{equation}
with $g=\frac{\eta-1}{\gamma}$, $d=\frac{\sqrt{M}b}{q\sqrt{N}}$, and $d_o=\frac{q}{\left(\sqrt{NM}b-q\left(1-N\right)\right)}$. To compute the largest eigenvalue $\lambda_{max}$ of $\mathbb{E}[\bm\Phi]$, we first write 
\begin{equation}
    \mathbb{E}[\bm\Phi]-\lambda\mathds{1}=(gd-\lambda-gd_0)\mathds{1}+\bm{g}\bm{d}_o^T \ ,
\end{equation}
where $\bm g = (g, \ldots, g)^T$ and $\bm d_0 = (d_0, \ldots, d_0)^T$. We can apply the matrix determinant lemma \cite{Harville1998MatrixAF} to give:
\begin{equation}
    \text{det}(\mathbb{E}[\bm\Phi]-\mathds{1}\lambda)=\left(1+\frac{Ngd_o}{g(d-d_o)-\lambda}\right)(g(d-d_o)-\lambda)^N.\label{determinant formula}
\end{equation}
We note from \eqref{determinant formula} that since $Ngd_o>0$ then $\mathbb{E}[\bm\Phi]$ has $N-1$ degenerate eigenvalues given by $g(d-d_o)$ and one large eigenvalue given by
\begin{equation}\label{e-value master}
    \tilde\lambda_{max}=g(d+d_o(N-1))\approx\frac{\eta-1}{\gamma}\left(\frac{b}{q\alpha}+\frac{q}{\frac{b}{\alpha}+q}\right)\ ,
\end{equation} 
in the large $N,M$ limit with fixed structure parameter $\alpha=\sqrt{\frac{N}{M}}$. This completes the proof of Eq. \eqref{lambda max master} in the main text.
\newpage

\section{Approximation used in Method \#3}\label{estimate calculations}
In order to arrive at the approximation used in Section \ref{Complex Math}, specifically Equation \eqref{approximated Phi}, we recall that we need to compute the constant
\begin{equation}\label{constant appendix}
    c=\frac{\mathbb{E}[W_{ij}]}{\mathbb{E}[X_{ij}]}\ ,
\end{equation}
where
\begin{equation}\label{X}
    \mathbb{E}[X_{ij}]=\frac{q}{\sqrt{NM}}\ .
\end{equation}
As for $\mathbb{E}[W_{ij}]$, we note that
\begin{align}
    \mathbb{E}[W_{ij}]&=\mathbb{E}\left[\frac{X_{ij}}{\sum_l^NX_{lj}}\right]
    =\int_0^\infty d\tau\mathbb{E}\left[X_{ij}e^{-\tau X_{ij}}\right]\prod_{l\neq i}\mathbb{E}\left[e^{-\tau X_{lj}}\right]\ ,
\end{align}
where we used the identity $1/x=\int_0^\infty d\tau e^{-\tau x}$ for $x>0$, and factorized the averages due to the independence of $X_{ij}$. Using $P(x)$ from Equation \eqref{eq:X distribution} we get
\begin{equation}\label{W estimate}
    \mathbb{E}[W_{ij}]=\int_0^\infty d\tau \int dx~P\left(x\right)x~e^{-\tau x}\left[\int dx~ P\left(x\right)e^{-\tau x}\right]^{N-1} \ .
\end{equation}
Now we also note that we can write
\begin{equation}
    \frac{d}{d\tau}\left[\int dx~ P\left(x\right)e^{-\tau x}\right]^{N}=-N\int dx~ P\left(x\right)x~e^{-\tau x}\left[\int dx P\left(x\right)e^{-\tau x}\right]^{N-1} \ .
\end{equation}
Hence
\begin{equation}
    \mathbb{E}[W_{ij}]=-\frac{1}{N}\int _0^\infty d\tau \frac{d}{d\tau}\left[\int dx P\left(x\right)e^{-\tau x}\right]^{N} \ .
\end{equation}
Recalling Equation \eqref{eq:X distribution}, we can therefore write
\begin{equation}
    \mathbb{E}[W_{ij}]=-\frac{1}{N}\left[\left(1-\frac{q}{\sqrt{NM}}\right)+\frac{q}{\sqrt{NM}}\mathbb{E}_K\left[e^{-\tau K}\right]\right]^N\Bigg|_{0}^\infty=-\frac{1}{N}\left[\left(1-\frac{q}{\sqrt{NM}}\right)^N-1\right] \ ,
\end{equation}
where we note that for strictly positive weights (such as those we are using), the weight dependence drops out. Hence in the limit $N,M\rightarrow\infty$ with fixed $\alpha=\sqrt{\frac{N}{M}}$, we have
\begin{equation}
    \mathbb{E}[W_{ij}]\simeq -\frac{1}{N}\left(e^{-q\alpha}-1\right) \ ,
\end{equation}
leading to 
%Thus inserting equations \eqref{constant} and \eqref{X}, we get the expression for our approximation constant:
\begin{equation}
    c=\frac{1-e^{-\alpha q}}{\alpha q}\ .
\end{equation}

\end{document}